\documentclass[acmsmall]{acmart}

\AtBeginDocument{%
  }

\usepackage{booktabs}    
\usepackage{longtable}
\usepackage{threeparttable}
\usepackage{array}
\usepackage{multirow}

\usepackage{cleveref}
\usepackage{colortbl}
\definecolor{DataCentric}{RGB}{237,125,49}
\definecolor{ModelingAlgo}{RGB}{68,114,196}
\definecolor{EvalExperiment}{RGB}{112,173,71}
\definecolor{PrivacySec}{RGB}{192,0,0}
\usepackage{subcaption}
\definecolor{DataCentric}{HTML}{E3F2FD}     % Light Blue
\definecolor{ModelingAlgo}{HTML}{FFF9C4}    % Pale Yellow
\definecolor{EvalExperiment}{HTML}{D1F2EB}  % Light Green
\definecolor{PrivacySec}{HTML}{FDEDEC}

\begin{document}
\title{A Comprehensive Review on Harnessing Large Language Models to Overcome Recommender System Challenges}
\author{Rahul Raja}
\affiliation{%
  \institution{Linkedin, Carnegie Mellon University, Stanford University}
  \country{USA}
}
% \email{rahul.1103}

\author{Anshaj Vats}
\affiliation{%
  \institution{Meta}
  % \city{Sunnyvale}
  % \state{California}
  \country{USA}
}
% \email{}

\author{Arpita Vats}
\affiliation{%
  \institution{Linkedin, Meta AI, Amazon, Boston University}
  % \city{Sunnyvale}
  % \state{California}
  \country{USA}
}
% \email{arpita@example.com}

\author{Anirban Majumder}
\affiliation{%
  \institution{Amazon}
  % \city{Bangalore}
  \country{USA}
}
% \email{anirban@example.org}
\thanks{This work does not relate to position at Linkedin, Meta, Amazon.}

\begin{abstract}
Recommender systems have traditionally followed modular architectures comprising candidate generation, multi-stage ranking, and re-ranking, each trained separately with supervised objectives and hand-engineered features. While effective in many domains, such systems face persistent challenges including sparse and noisy interaction data, cold-start problems, limited personalization depth, and inadequate semantic understanding of user and item content. The recent emergence of Large Language Models (LLMs) offers a new paradigm for addressing these limitations through unified, language-native mechanisms that can generalize across tasks, domains, and modalities. In this paper, we present a comprehensive technical survey of how LLMs can be leveraged to tackle key challenges in modern recommender systems. We examine the use of LLMs for prompt-driven candidate retrieval, language-native ranking, retrieval-augmented generation (RAG), and conversational recommendation, illustrating how these approaches enhance personalization, semantic alignment, and interpretability without requiring extensive task-specific supervision. LLMs further enable zero- and few-shot reasoning, allowing systems to operate effectively in cold-start and long-tail scenarios by leveraging external knowledge and contextual cues. We categorize these emerging LLM-driven architectures and analyze their effectiveness in mitigating core bottlenecks of conventional pipelines. In doing so, we provide a structured framework for understanding the design space of LLM-enhanced recommenders, and outline the trade-offs between accuracy, scalability, and real-time performance. Our goal is to demonstrate that LLMs are not merely auxiliary components but foundational enablers for building more adaptive, semantically rich, and user-centric recommender systems

% In this paper, we provide a comprehensive technical survey and design perspective on how LLMs are transforming recommender systems. We characterize the evolution from two-tower retrieval and cascade architectures to LLM-augmented pipelines that support prompt-based scoring, retrieval-augmented generation (RAG), and language-native explanation mechanisms. We analyze architectural trade-offs involving latency, generalization, interpretability, and personalization depth across LLM integration points. Additionally, we identify limitations related to scalability, prompt drift, privacy, and evaluation robustness, and discuss open challenges including alignment, observability, and hybridization with traditional models. Our work provides a structured foundation for understanding the design space of LLM-enhanced recommenders and highlights future research directions at the intersection of language modeling and personalized ranking.
\end{abstract}
\maketitle
% \vspace{-0.3em}

\section{Introduction}
Recommender systems have become essential across a broad spectrum of digital applications, including content streaming, e-commerce, education, recruiting, and social media platforms~\cite{gomez2022netflix, cheng2023amazon, linkedin2019twotower}. From personalized playlists on Spotify to tailored learning paths in MOOCs and targeted ads on LinkedIn, recommender systems play a pivotal role in shaping user experience and driving engagement. Despite their widespread adoption and continuous improvements, these systems face enduring challenges such as data sparsity, cold-start problems, dynamic user interests, and explainability~\cite{panigrahi2023recosys}. 
As modern pipelines grow in complexity—with multi-stage architectures, large-scale retrieval, and diverse modalities—the demands for scalability, transparency, and adaptability have intensified. Recommender systems have traditionally followed modular architectures comprising candidate generation, multi-stage ranking, and re-ranking—each trained separately using supervised objectives and hand-engineered features ~\citep{Chen_2023}. While effective in many domains, such systems face persistent challenges, including sparse and noisy interaction data, cold-start problems, limited personalization depth, and inadequate semantic understanding of user and item content.\\
The recent emergence of Large Language Models (LLMs) offers a new paradigm for addressing these limitations through unified, language-native mechanisms that can generalize across tasks, domains, and modalities. In this paper, we present a comprehensive technical survey of how LLMs can be leveraged to tackle key challenges in modern recommender systems. We examine the use of LLMs for prompt-driven candidate retrieval, language-native ranking, retrieval-augmented generation (RAG), and conversational recommendation—illustrating how these approaches enhance personalization, semantic alignment, and interpretability without requiring extensive task-specific supervision. LLMs also enable zero- and few-shot reasoning, allowing systems to operate effectively in cold-start and long-tail scenarios by leveraging external knowledge and contextual cues. We categorize these emerging LLM-driven architectures and analyze their effectiveness in mitigating core bottlenecks of conventional pipelines. In doing so, we provide a structured framework for understanding the design space of LLM-enhanced recommender systems, and outline trade-offs between accuracy, scalability, and real-time performance.\\
Our work provides a foundation for researchers and engineers to reimagine recommender systems in the era of large language modeling, and highlights promising directions for future innovation.

\begin{figure}[h!]  % Use [htbp] if you don't want it fixed here
    \centering
    \includegraphics[width=1.0\textwidth]{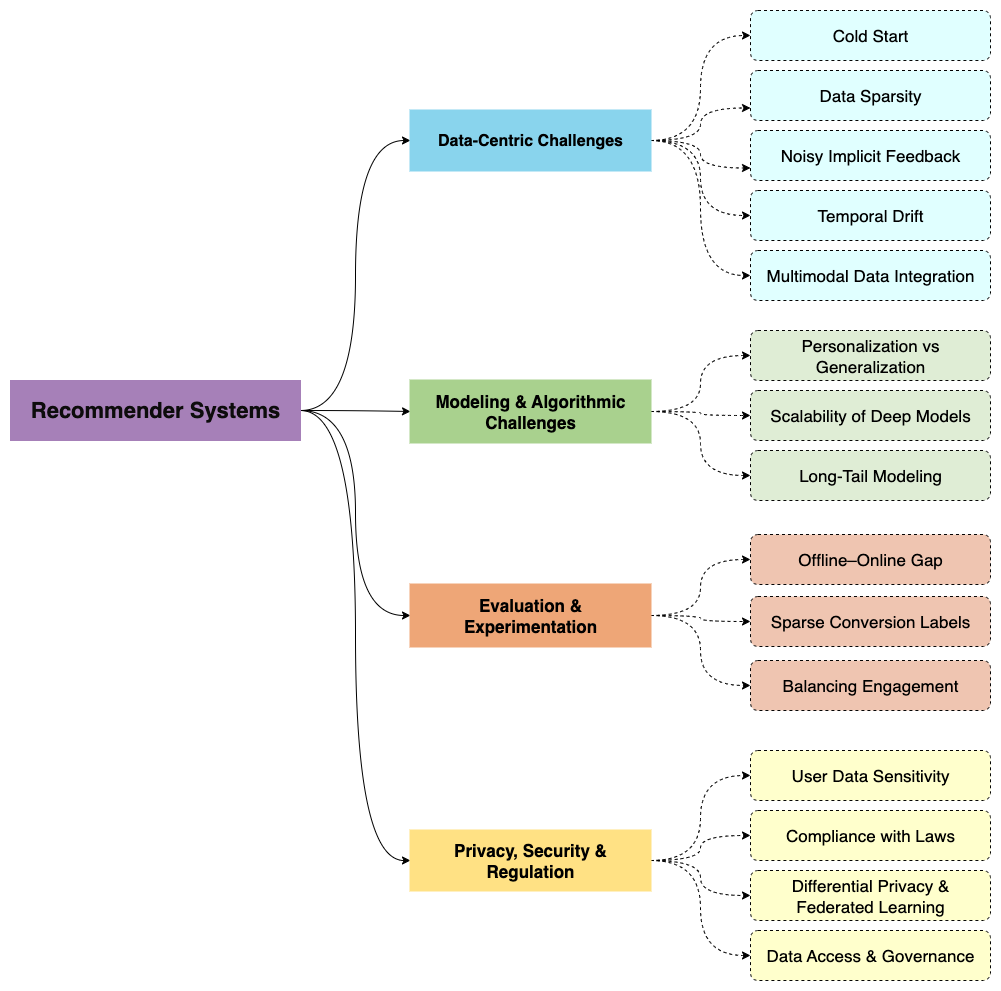}  % Adjust width as needed
    \caption{Taxonomy of recommender system challenges and corresponding LLM-based solutions. The framework categorizes challenges into four major areas—data-centric, modeling and algorithmic, evaluation, and privacy/security—each with sub-problems and LLM-driven strategies for addressing them.'=}
    \label{fig:componenets_recommendation}
\end{figure}

\section{Evolution of Architectures in Recommender Systems}
Recommender systems have progressed from heuristic rules to collaborative filtering, latent factorization, neural and graph-based models, and finally industrial-scale retrieval frameworks. Each paradigm enhanced expressiveness, scalability, and the ability to capture complex dependencies.

\subsection{From Heuristics to Latent Factor Models}
Early recommenders used deterministic heuristics such as popularity- and co-occurrence-based rules~\cite{schafer2001commerce, bobadilla2013recommender, sarwar2001itembased}, which scale well but lack personalization. Collaborative Filtering (CF)~\cite{7176109} introduced personalization by exploiting similarity in interaction patterns between users or items. While simple and interpretable, CF suffers from sparsity and cold-start issues.  

Matrix Factorization (MF)~\cite{1102314, koren2009matrix} addressed scalability by embedding users and items into a shared low-dimensional space, with predictions computed via inner products of latent vectors. Variants such as ALS~\cite{zhou2008large} and SVD++~\cite{koren2009matrix} incorporated implicit feedback and became widely adopted in industrial systems due to their balance of accuracy and efficiency.

\subsection{Hybrid and Neural Extensions}
Content-Based Filtering (CBF)~\cite{lops2011content} enriched recommendations with item attributes such as textual, categorical, or visual embeddings, making it more robust to cold-start users. However, CBF tends to overspecialize. Hybrid models~\cite{burke2002hybrid} combined collaborative and content signals, exemplified by LightFM~\cite{kula2015metadata} and DeepFM~\cite{guo2017deepfm}, which integrate metadata with learned embeddings.  

Deep learning extended this line with Neural Collaborative Filtering (NCF)~\cite{he2017neuralcollaborativefiltering}, which replaces linear interaction functions with nonlinear neural networks, and Wide \& Deep~\cite{cheng2016wide}, which jointly optimizes memorization and generalization. These architectures established the basis for modern industrial recommenders.

\subsection{Sequential and Graph-Based Models}
User behavior often exhibits temporal and contextual dynamics. Sequence-aware recommenders such as GRU4Rec~\cite{hidasi2015session} applied recurrent neural networks to session data, while transformer-based approaches (e.g., SASRec~\cite{kang2018self}, BERT4Rec~\cite{sun2019bert4rec}, TiSASRec~\cite{li2020time}) leveraged self-attention to capture long-range dependencies and temporal intervals. These models outperform traditional CF in sequential domains such as streaming and e-commerce.  

In parallel, graph-based recommenders~\cite{wang2021graphlearningbasedrecommender} represented interactions as bipartite graphs. Neural Graph Collaborative Filtering (NGCF)~\cite{ying2018graph} propagated embeddings through nonlinear message passing, while LightGCN~\cite{he2020lightgcn} simplified this to linear aggregation, improving both accuracy and scalability. PinSage~\cite{ying2018graph} scaled graph-based recommendation to billions of nodes using random walks and sampling, enabling deployment in web-scale platforms like Pinterest.

\subsection{Industrial-Scale Retrieval Frameworks}
With item catalogs often exceeding billions, retrieval efficiency became a dominant concern. Two-tower architectures~\cite{yuan2024contextgnntwotowerrecommendationsystems} learn independent embeddings for users and items, with relevance computed as the inner product between vectors. Pre-computed item embeddings can be indexed using Approximate Nearest Neighbor (ANN) search, enabling sublinear retrieval times.  

The YouTube DNN~\cite{covington2016deep} demonstrated the effectiveness of deep two-tower retrieval with sampled softmax, while Facebook’s DLRM~\cite{naumov2019deep} integrated sparse and dense features with cross-feature interaction layers. These retrieval modules form the candidate generation stage in multi-stage pipelines, with re-ranking typically handled by more expressive models such as gradient boosting or transformers.

\section{Key Components of Recommender System Pipelines}
% \begin{figure}[H]  % Use [htbp] if you don't want it fixed here
%     \centering
%     \includegraphics[width=1.0\textwidth]{Key_Componenets.png}  % Adjust width as needed
%     \caption{The diagram illustrates a typical three-stage architecture in modern recommender systems: (1) Candidate Generation, which retrieves a broad set of potentially relevant items; (2) Scoring and Ranking, which computes personalized relevance scores and sorts candidates using deep models; and (3) Post-Ranking and Calibration, which adjusts the ranked list for diversity, fairness, freshness, or exploration. Each stage progressively refines the item list to optimize user satisfaction and system-level goals.}
%     \label{fig:componenets_recommendation}
% \end{figure}
\subsection{Candidate Generation}
Candidate generation is the first stage in large-scale recommender systems, responsible for filtering a massive corpus of items down to a smaller subset of potentially relevant candidates for a given user. This stage prioritizes recall over precision, ensuring that relevant items are included in the candidate set, even if some irrelevant items are also retrieved. Precision is handled in later stages through more computationally intensive ranking models. The goal is to efficiently retrieve a set of items \( \mathcal{C}(u) \subset \mathcal{I} \) for a user \( u \in \mathcal{U} \), where \( \mathcal{I} \) denotes the entire item corpus. Formally, candidate generation can be defined as retrieving all items \( i \in \mathcal{I} \) such that the similarity between the user embedding \( f_u \) and the item embedding \( f_i \) exceeds a threshold \( \tau \), i.e.,

\[
\mathcal{C}(u) = \{ i \in \mathcal{I} \mid \text{sim}(f_u, f_i) \geq \tau \},
\]

where \( \text{sim}(\cdot, \cdot) \) is a similarity function such as dot product or cosine similarity. Several techniques are commonly employed in candidate generation. Approximate Nearest Neighbor (ANN) search allows for efficient similarity-based retrieval using indexing structures like HNSW or product quantization~\cite{malkov2018efficient, jegou2011product}. Graph-based methods leverage user-item interaction graphs, using algorithms such as random walks~\cite{perozzi2014deepwalk} or personalized PageRank~\cite{haveliwala2002topic} to identify semantically or behaviorally related items. Heuristic filtering approaches use hand-crafted rules based on metadata, popularity, or recent activity to shortlist items quickly. Embedding-based retrieval methods, often built using two-tower or dual encoder models, learn representations for users and items such that similar entities lie close in a shared vector space~\cite{yang2020mixed}. Additionally, session-based models take into account recent user interactions and temporal patterns, often utilizing RNNs~\cite{hidasi2015session} or attention mechanisms~\cite{li2017neural} to generate context-aware candidates.

Effective candidate generation requires balancing computational efficiency with the need for broad item coverage. Systems must be optimized for low latency and high throughput while ensuring that the candidate pool includes a diverse and relevant set of items to support personalization downstream. Periodic updates to item indices and embeddings are essential to maintain temporal relevance. Overall, the success of downstream ranking models is closely tied to the quality and coverage of the candidate set produced in this stage.

\subsection{Scoring and Ranking}
Once the candidate set is generated, the next stage in the recommendation pipeline involves computing a relevance score for each candidate and producing a ranked list. This scoring process is handled by machine learning models that are typically more expressive and computationally expensive than those used in candidate generation, since they operate on a smaller set of items. These models ingest high-dimensional features encompassing user profiles, item attributes, and contextual signals such as time of day, device type, and recent interactions. Common architectures for this stage include gradient-boosted decision trees (GBDT)~\cite{ke2017lightgbm}, deep neural networks (DNNs)~\cite{covington2016deep}, and transformer-based models~\cite{sun2019bert4rec}, which are trained to predict item relevance based on learned representations of user-item pairs.

Formally, for each user \( u \in \mathcal{U} \) and candidate item \( i \in \mathcal{C}(u) \), the system learns a scoring function \( s(u, i) \) defined as:

\[
s(u, i) = f_\theta(\phi_u, \phi_i, \psi_{u,i}),
\]

where \( \phi_u \) and \( \phi_i \) are the embedding representations of the user and item, respectively, and \( \psi_{u,i} \) denotes additional features capturing user-item interaction context. The function \( f_\theta \) is typically a feed-forward neural network or transformer-based encoder that outputs a scalar relevance score. The top-ranked items are selected by sorting candidates in descending order of \( s(u, i) \).

In multi-task learning settings, the model predicts several outcomes simultaneously—such as click probability, dwell time, and engagement—using a shared encoder and task-specific output heads:

\[
h = \text{Encoder}_\theta(\phi_u, \phi_i, \psi_{u,i}),
\]

\[
\hat{y}_{\text{click}} = \sigma(W_{\text{click}} h), \quad
\hat{y}_{\text{dwell}} = \text{ReLU}(W_{\text{dwell}} h), \quad
\hat{y}_{\text{engage}} = \sigma(W_{\text{engage}} h),
\]

where \( h \) is the latent representation, \( \sigma \) denotes the sigmoid function, and \( W \)'s are task-specific projection matrices. The total training loss is computed as a weighted combination of individual task losses:

\[
\mathcal{L} = \lambda_1 \cdot \mathcal{L}_{\text{click}} + \lambda_2 \cdot \mathcal{L}_{\text{dwell}} + \lambda_3 \cdot \mathcal{L}_{\text{engage}}.
\]

This setup enables joint optimization across objectives and improves the model's ability to learn shared behavioral patterns. Domain-specific features further enhance personalization, such as temporal trends, item freshness, category information, and content embeddings. The scoring and ranking stage is critical for final recommendation quality, as it ultimately determines the ordering and presentation of content to the user.
% \begin{table}[ht]
% \centering
% \resizebox{\textwidth}{!}{%
% \begin{tabular}{l|c|c|c|c}
% \toprule
% \textbf{Reranker Type} & \textbf{Latency (ms)} & \textbf{Generalization} & \textbf{Explainability} & \textbf{Personalization Depth} \\
% \midrule
% MLP/DNN Reranker & $\sim$5--10 & Low (in-domain only) & Limited & Medium \\
% BERT-based Reranker & $\sim$30--70 & Medium & Moderate & High \\
% Prompt-based LLM & $\sim$300+ & High (zero-shot) & High & High \\
% LLM+Embedding Fusion & $\sim$100--150 & Medium--High & Moderate & Very High \\
% \bottomrule
% \end{tabular}
% }
% \caption{
% \textbf{Latency and Capabilities of Reranker Architectures.}
% Prompt-driven LLMs provide deeper semantic personalization and explainability, but incur high latency. Hybrid strategies offer better trade-offs under latency constraints.
% }
% \label{tab:reranker_comparison}
% \end{table}

\subsection{Post-Ranking and Calibration}
After the initial scoring and ranking phase, post-ranking modules are applied to adjust the ranked list in order to align recommendations with broader system-level goals beyond user-item relevance. These objectives include promoting content diversity, controlling item fatigue, ensuring fairness across item categories or creators, and maintaining freshness in temporal contexts. This stage is critical to shaping the final user experience and achieving long-term business metrics such as content exposure, retention, and platform health.

The input to this module is an already ranked list \( \mathcal{R}(u) = \{ i_1, i_2, \dots, i_k \} \) with associated relevance scores \( s(u, i_j) \). Post-ranking adjusts these scores via a transformation function \( g(\cdot) \), which considers additional factors such as diversity penalties or freshness boosts:

\[
s'(u, i_j) = g(s(u, i_j), \delta(i_j), \gamma(i_j)),
\]

where:
\begin{itemize}
    \item \( s(u, i_j) \) is the original relevance score,
    \item \( \delta(i_j) \) is a diversity-aware adjustment factor (e.g., based on item similarity or content clusters),
    \item \( \gamma(i_j) \) is a freshness or recency factor.
\end{itemize}

One common technique is \textbf{diversity-aware re-ranking}, which penalizes redundant content in the top positions using a marginal gain function. This can be formalized using Determinantal Point Processes (DPP)~\cite{kulesza2012determinantal} or greedy submodular optimization~\cite{ashkan2015optimal}. Similarly, \textbf{fatigue control} mechanisms reduce the score of items already seen or interacted with recently, preventing oversaturation. 

In dynamic settings, \textbf{exploration-exploitation trade-offs} are addressed through bandit algorithms~\cite{li2010contextual} or reinforcement learning~\cite{chen2019topk}. These methods adjust the final ranking to allow new or uncertain items to be exposed to users with non-zero probability. In a contextual bandit framework, the post-ranking score \( s' \) may incorporate expected reward:

\[
s'(u, i_j) = \mathbb{E}[r_{u, i_j} \mid x_{u,i_j}],
\]

where \( r_{u, i_j} \) is the reward (e.g., click, engagement) and \( x_{u,i_j} \) are the contextual features of the interaction.

Overall, post-ranking ensures that the final list achieves a desirable trade-off between individual relevance and system-level goals. This stage plays a critical role in aligning model outputs with platform constraints and policy requirements while enhancing long-term user satisfaction.

\section{Evolving Recommender Systems: Addressing Traditional Challenges with LLM-Based Solutions }
Recommender systems deployed in industrial settings face a wide array of challenges, many of which stem from the nature of the data itself. Below, we outline key data-centric issues and describe how leading platforms address them using scalable machine learning techniques, representation learning, and multimodal data fusion.

In this section, we explore how LLMs can be integrated into Recommender system architectures to address six main categories of industrial challenges. We begin with data understanding and cold start issues and then progress through modeling, evaluation, system design, privacy, and organizational concerns. For each area, we discuss not only the strengths of LLM-based approaches but also the practical trade-offs and open research directions they entail.

\begin{center}
\scriptsize
\begin{longtable}{p{1.8cm}|p{2cm}|p{4cm}|p{2.8cm}}
  \caption{Challenges in Recommender Systems and Corresponding LLM-Based Solutions}
  \label{tbl:llm_challenges} \\
  \toprule
  \textbf{Challenges} & \textbf{Sub-problem}
    & \textbf{LLM-Based Solutions} & \textbf{References} \\
  \midrule
  \endfirsthead
  
  \multicolumn{4}{c}%
  {{\bfseries \tablename\ \thetable{} -- continued from previous page}} \\
  \toprule
  \textbf{Challenges} & \textbf{Sub-problem}
    & \textbf{LLM-Based Solutions} & \textbf{References} \\
  \midrule
  \endhead
  
  \midrule \multicolumn{4}{r}{{Continued on next page}} \\
  \endfoot
  
  \bottomrule
  \endlastfoot

  % ---------- DATA-CENTRIC --------------------------------------
  \rowcolor{DataCentric}
  \multirow{5}{1.8cm}{\textbf{Data-Centric}}
  & Cold Start &
    Content-Conditioned Generation \newline
    Retrieval-Augmented Generation \newline
    Zero-Shot Personalization \newline
    Representation Bootstrapping \newline
    Language-Native Dialogue \newline
    Prompt-Based Conditioning \newline
    Multimodal Embedding Synthesis \newline
    Meta-Learning for Fast Adaptation \newline
    Cross-Domain Transfer &
    \citep{zhang2021language, li2023generative, jin2023sam, xu2023llmr, kang2023llmrec, dai2023promptrec} \newline
    \citep{huang2024surveyretrievalaugmentedtextgeneration, gao2021rllmrec, perez2021, asai2020learning} \newline
    \citep{wei2022finetunedlanguagemodelszeroshot, purificato2024usermodelinguserprofiling, thomas2022spotify} \newline
    \citep{tennenholtz2024demystifyingembeddingspacesusing, huber2025embeddingtoprefixparameterefficientpersonalizationpretrained} \newline
    \citep{chan2024humanllmbasedvoiceassistant, wang2024dialogrec, zheng2024unifiedparadigmintegratingrecommendation} \newline
    \citep{ramos2025peapodpersonalizedpromptdistillation, jiang2024prompttuningitemcoldstart} \newline
    \citep{gadre2023video, alayrac2022flamingo, hou2023mmrec, yu2022modality} \newline
    \citep{wang2024limamlpersonalizationdeeprecommender, finn2017maml, lee2019melu}\newline
    \citep{vajjala2024crossdomainrecommendationmeetslarge, yi2023contrastivegraphprompttuningcrossdomain, Liu_2025,vats2024exploringimpactlargelanguage}\\
  \cmidrule{2-4}
  \rowcolor{DataCentric}
  & Data Sparsity &
    Text-Driven Generalization \newline
    Semantic Matching via Embedding &
    \citep{zhang2021b, cui2023genrecs, chen2023prompt} \newline
    \citep{wang2023llm4coldstart, cui2024promptbasedknowledgegraphfoundation}\\
  \cmidrule{2-4}
  \rowcolor{DataCentric}
  & Noisy Implicit Feedback &
    Feedback Interpretation with Contextual Prompts &
    \citep{bodonhelyi2024userintentrecognitionsatisfaction, zhang2023lightrec, lee2013pseudo, hu2016label, jin2023sam}\\
  \cmidrule{2-4}
  \rowcolor{DataCentric}
  & Temporal Drift &
    Temporal Adaptation with LLMs \newline
    Structural and Hybrid Integration &
    \citep{zhao2025videoexpertaugmentedllmtemporalsensitive, kumar2025leveragingknowledgegraphsllms} \newline
    \citep{luo2024integratinglargelanguagemodels, han2025rethinkingllmbasedrecommendationsquery}\\
  \cmidrule{2-4}
  \rowcolor{DataCentric}
  & Multimodal Data Integration &
    Modal-Aware Weighting \newline
    Unified Representation \newline
    Cross-Modal Alignment \newline
    Multimodal Imputation \newline
    Semantic Fusion &
    \citep{xu2023llmr, jin2023sam, ma2024xrec} \newline
    \citep{lu2022unified, lu2023unifiedio2scalingautoregressive} \newline
    \citep{sapkota2025multimodallargelanguagemodels, zhang2025unifiedmultimodalunderstandinggeneration} \newline
    \citep{radford2021learning, zhou2025omnifm, chen2023pali} \newline
    \citep{jung2025learninggeneralizablepromptclip, aghajanyan2022cm3causalmaskedmultimodal}\\
  \midrule

  % ---------- MODELING & ALGORITHMIC ----------------------------
  \rowcolor {ModelingAlgo}
  \multirow{3}{1.8cm}{\textbf{Modeling \& Algorithmic}}
  & Personalization vs. Generalization &
    Instruction-Tuned Generalization \newline
    Prompt-Tuned Personalization \newline
    Behavioral Diversity \newline
    Multitask Prompting &
    \citep{liu2025surveypersonalizedlargelanguage} \newline
    \citep{jiang2024prompttuningitemcoldstart, li2022pepler} \newline
    \citep{wang2023instrec}\\
  \cmidrule{2-4}
  \rowcolor {ModelingAlgo}
  & Scalability of Deep Models &
    LLM-Based Distillation \newline
    Prompt-Efficient Inference \newline
    Two-Stage Hybrid Pipelines \newline
    Sparse Activation Architectures &
    \citep{chung2022scaling, jiang2024prompttuningitemcoldstart, wang2025rdrecrationaledistillationllmbased} \newline
    \citep{Zhao2022, shen2021powerbert, li2023generative, chen2023llm} \newline
    \citep{hu2021lora, lester2021power, jin2023sam, ouyang2022instructgpt} \newline
    \citep{du2022glam, lin2021m6t}\\
  \cmidrule{2-4}
  \rowcolor {ModelingAlgo}
  & Long-Tail Modeling &
    Content-Enriched Generation \newline
    Retrieval-Augmented Tail Expansion \newline
    Tail-Aware Few-Shot Prompting \newline
    Multimodal Tail Representation &
    \citep{zhang2021language, kang2023llmrec} \newline
    \citep{vasile2022, dai2023promptrec, wu2024coralcollaborativeretrievalaugmentedlarge} \newline
    \citep{chung2022scaling, wang2023instrec} \newline
    \citep{vasile2022, oh2025understandingmultimodalllmsdistribution}\\
  \midrule

  % ---------- EVALUATION & EXPERIMENTATION ----------------------
  \rowcolor{EvalExperiment}
  \multirow{3}{1.8cm}{\textbf{Evaluation \& Experimentation}}
  & Offline–Online Gap &
    Counterfactual Evaluation \newline
    Offline Metric Recalibration \newline
    Behavior-Level Satisfaction \newline
    Interactive User Simulation \newline
    Evaluation Metric Generation &
    \citep{gilotte2018offline, saito2020unbiased, jagerman2022evaluating} \newline
    \citep{chen2023opportunitieschallengesofflinereinforcement, wu2024surveylargelanguagemodels} \newline
    \citep{zhao2025llmsrecognizepreferencesevaluating, lee2023satisfaction} \newline
    \citep{shang2025agentrecbenchbenchmarkingllmagentbased, bemthuis2024crispdmbasedmethodologyassessingagentbased} \newline
    \citep{zhou2023llmjudge, alikhani2023gptmetrics, xu2023llmevaltemplate}\\
  \cmidrule{2-4}
  \rowcolor{EvalExperiment}
  & Sparse Conversion Labels &
    Proxy Signal Augmentation \newline
    Instruction-Tuned Label Imputation \newline
    Generative Multi-Task Learning \newline
    Counterfactual Label Reasoning \newline
    Language-Guided Reweighting &
    \citep{dai2023promptrec, jin2023sam, zhang2021language} \newline
    \citep{wong2025highfidelitypseudolabelgenerationlarge} \newline
    \citep{freiberger2025prismenovelllmpoweredtool} \newline
    \citep{joshi2024llmspronefallaciescausal} \newline
    \citep{tang2023recent}\\
  \cmidrule{2-4}
  \rowcolor{EvalExperiment}
  & Balancing Immediate Engagement &
    LLM-Based Proxy Reward Estimation \newline
    Counterfactual Dialogue \newline
    Preference Drift Detection \newline
    Multi-Objective RL with LLM-Guided Reward &
    \citep{xue2022rlhfrec, park2019clicksat} \newline
    \citep{chen2021intent, li2022crslab} \newline
    \citep{zheng2023usergptsim} \newline
    \citep{zhao2024stablerec, lee2023temprecmemory} \newline
    \citep{christiano2017deep, liu2023aligning, ziegler2019fine}\\
  \midrule
  
  % ---------- PRIVACY, SECURITY & REGULATORY --------------------
  \rowcolor{PrivacySec}
  \multirow{4}{1.8cm}{\textbf{Privacy, Security, \& Regulatory}}
  & User Data Sensitivity &
    Token Attribution and Prompt Fingerprinting &
    \citep{gupta2024promptdrift}\\
  \cmidrule{2-4}
  \rowcolor{PrivacySec}
  & Compliance with Laws &
    Streaming Log Replay + Token-Level Attribution &
    \citep{xie2023, park2025blackbox}\\
  \cmidrule{2-4}
  \rowcolor{PrivacySec}
  & Differential Privacy \& Federated Learning &
    LLMs for Ephemeral Personalization \newline
    Gradient-Sanitized Fine-Tuning \newline
    Federated Prompting &
    \citep{mcmahan2017fl} \newline
    \citep{nguyen2024ephemeral, rao2024synthetic} \newline
    \citep{li2025dpgradient, zhou2025dpgen, wu2025federated}\\
  \cmidrule{2-4}
  \rowcolor{PrivacySec}
  & Data Access and Governance &
    Data Minimization via LLMs \newline
    Instructable Privacy Filters \newline
    Synthetic Data Generation &
    \citep{bougie2025simusersimulatinguserbehavior, data_governance_examples} \newline
    \citep{chen2025vlmguardr1proactivesafetyalignment, dong2024safeguardinglargelanguagemodels} \newline
    \citep{xie2024differentiallyprivatesyntheticdata, lin2024surveydiffusionmodelsrecommender}\\
  
\end{longtable}
\end{center}

\subsection{Data-Centric Challenges}
Recommender systems deployed in industrial environments face a host of challenges rooted not just in algorithmic design, but in the quality, availability, and dynamics of the underlying data. The data-centric issues significantly influence model generalization, fairness, personalization, and system robustness. Below, we dissect key categories of data-centric challenges and discuss how they impact large-scale deployment and learning dynamics in recommender pipelines.

\subsubsection{Cold Start Problem}

The \textit{cold start problem} refers to the difficulty of generating reliable recommendations for new users or new items due to the absence of interaction history. This issue is particularly acute in collaborative filtering (CF) systems, where latent representations are learned from observed user–item interactions.

In matrix factorization frameworks, predicted preferences are typically computed as:
\[
\hat{r}_{ui} = \mathbf{p}_u^\top \mathbf{q}_i
\]
where $\mathbf{p}_u, \mathbf{q}_i \in \mathbb{R}^d$ are the latent embeddings of user $u$ and item $i$. These embeddings are optimized by minimizing a loss over the set of observed interactions $\mathcal{O}$:
\[
\min_{\{\mathbf{p}_u\}, \{\mathbf{q}_i\}} \sum_{(u,i)\in \mathcal{O}} \mathcal{L}(\hat{r}_{ui}, r_{ui}) + \lambda \left( \|\mathbf{p}_u\|^2 + \|\mathbf{q}_i\|^2 \right)
\]
In cold start settings, when $u' \notin \mathcal{O}_u$ or $i' \notin \mathcal{O}_i$, the corresponding embeddings $\mathbf{p}_{u'}$ or $\mathbf{q}_{i'}$ are poorly initialized or undefined, leading to degraded or invalid predictions:
\[
\hat{r}_{u'i} = \mathbf{p}_{u'}^\top \mathbf{q}_i \quad \text{or} \quad \hat{r}_{u i'} = \mathbf{p}_u^\top \mathbf{q}_{i'}
\]

This challenge is exacerbated in large-scale, dynamic environments where thousands of new users and items enter the system daily~\cite{quadrana2017sequence}. Even more complex architectures such as two-tower models~\cite{covington2016deep} or deep sequential recommenders~\cite{hidasi2015session} rely on embedding representations learned from prior interactions, and thus remain vulnerable to cold start scenarios. The sparsity of user--item signals and the lack of initialization paths for unseen entities make cold start a persistent bottleneck in the design of effective recommender systems~\cite{schein2002methods,park2006naive}. LLMs offer a paradigm shift in mitigating the cold start problem by leveraging their ability to process rich textual and contextual information without requiring historical interaction data~\cite{dai2023promptrec,wang2023llm4coldstart}. Unlike traditional collaborative filtering methods that depend on dense user--item matrices, LLMs can infer relevance through semantic reasoning, prompt-based generation, and cross-modal understanding. Let's examine how LLMs can solve the cold start problem in recommender systems.

\vspace{0.5em}
\textbf{Content-Conditioned Generation –}
LLMs pretrained on web-scale corpora possess rich world knowledge and linguistic priors, enabling them to generate item recommendations solely from textual descriptions. This paradigm is particularly advantageous in cold-start and low-resource scenarios where interaction histories are sparse or nonexistent. Traditional recommender systems often struggle in these settings, as they depend heavily on collaborative signals (e.g., clicks, ratings, watch time). In contrast, LLMs can interpret and reason over content attributes such as titles, descriptions, reviews, genres, and even product specifications.

For instance, Generative Recommenders~\cite{zhang2021language, li2023generative} employ prompt-based methods where user queries or profiles are combined with product metadata to produce ranked outputs or candidate lists. In these setups, the generation process bypasses the need for learned user-item embeddings and instead leverages natural language priors encoded in the LLM:

\[
\hat{r}_{ui} = \text{LLM}(\texttt{``User profile: ''} + x_u + \texttt{`` Item: ''} + x_i)
\]

Such models can be further enhanced with instruction prompts, prefix tuning, or adapters to align generation with domain-specific preferences and intents. Recent work demonstrates that even simple prompts can elicit meaningful ranking signals, especially when the model is instructed to output preference scores, reasoning traces, or direct answers.

This approach has been adopted in real-world pipelines, including JD.com's product recommendation and Pinterest's item-to-item generation~\cite{jin2023sam, xu2023llmr}, where structured item metadata and textual attributes are encoded as input contexts. Furthermore, systems like GPTRec and PromptRec have shown how generative LLMs can synthesize interaction labels (e.g., likelihood of liking or purchasing an item) from textual descriptions, enabling bootstrap datasets for pretraining, self-training, or teacher-student distillation in downstream rankers.

By conditioning generation on content alone, these models enable rapid personalization without requiring prior interactions, thereby offering a scalable solution to cold-start recommendation, catalog expansion, long-tail item surfacing, and new-user onboarding. As instruction tuning and few-shot prompting mature, we expect content-conditioned generation to play a central role in building adaptable, zero-shot capable recommender agents that generalize across domains with minimal supervision.

\vspace{0.5em}
\textbf{Retrieval-Augmented Generation (RAG) –}  
RAG-based methods enhance LLMs with a retrieval module that selects relevant items or documents based on the input query. In recommender systems, this architecture decouples memorization from inference, allowing the model to attend to similar items, behavioral patterns, or metadata even when the target entity is entirely new or sparsely observed~\cite{kang2023llmrec, dai2023promptrec}. By combining parametric knowledge from the LLM with non-parametric access to a large external index, RAG frameworks enable more accurate, controllable, and interpretable recommendation outputs.

Formally, the LLM decoder conditions on the retrieved context $\mathcal{R}(x_u)$ to generate a recommendation or compute a relevance score:
\[
P(i \mid x_u) = \text{Decoder}_{\text{LLM}}(x_u, \mathcal{R}(x_u))
\]
where $\mathcal{R}(x_u)$ denotes a retrieved support set from a product, user, or content database, often implemented using approximate nearest neighbor (ANN) search or semantic retrievers (e.g., dual-encoder models or dense passage retrievers).

In practice, retrieval can be conditioned on user intent, content type, or temporal signals, enabling RAG pipelines to dynamically personalize the input context ~\cite{huang2024surveyretrievalaugmentedtextgeneration}. For instance, retrieved exemplars can include past purchases by similar users ~\cite{gao2021rllmrec}, product FAQs, reviews, or even click sequences from related users. The decoder then performs sequence-level reasoning over this context to either generate ranking scores, provide justifications (e.g., “because you liked X…”), or directly synthesize recommendations.

Amazon Alexa's RecRAG and OpenAI’s plugin-based recommendation prototypes have explored such architectures in production-like environments~\cite{perez2021, li2023generative}, demonstrating their applicability in real-time systems. These systems often include fast-refresh retrievers and pre-encoded knowledge bases to ensure responsiveness at inference time.

Recent advancements such as contrastive retrieval tuning, multi-hop retrieval~\cite{asai2020learning}, and instruction-based reranking~\cite{dong2024dontforgetconnectimproving} are further improving the efficacy of RAG for RecSys. Moreover, hybrid approaches that blend retrieval with fine-tuned rerankers or LLM feedback loops (e.g., re-querying or dynamic re-ranking) are emerging as state-of-the-art strategies for multi-objective personalization tasks ~\cite{gao2025llm4rerankllmbasedautorerankingframework}.
Looking ahead, RAG presents a promising direction for bridging symbolic reasoning (via retrieval) and generative modeling, allowing recommender systems to scale across domains, maintain explainability, and adapt to evolving content landscapes with minimal retraining~\cite{shi2023rag4rec}.

\vspace{0.5em}
\textbf{Zero-Shot Personalization –}  
Instruction-tuned LLMs (e.g., \textsc{GPT-4} ~\cite{openai2023gpt4}, \textsc{FLAN-T5} ~\cite{chung2022scaling}) are capable of zero-shot recommendation through natural language queries~\cite{wei2022finetunedlanguagemodelszeroshot}. These models are pretrained to follow open-ended instructions and can perform complex reasoning and generation tasks without explicit task-specific fine-tuning. In the recommendation context, they generalize to new users or items by leveraging semantic cues from textual metadata, behavior descriptions, or demographic features, bypassing the need for supervised training or interaction histories on the specific platform ~\cite{purificato2024usermodelinguserprofiling}.

This capability enables highly flexible recommendation systems where users can describe their preferences in natural language e.g., “I’m looking for a sci-fi show with philosophical themes like Black Mirror” or “Suggest books similar to what a data science graduate might enjoy.” The model can interpret this intent, perform semantic matching over a corpus of content metadata, and return high-quality, context-aware suggestions without explicit training data.

Spotify's conversational agents ~\cite{thomas2022spotify} and TikTok’s personalization assistant have applied these capabilities to enhance cold-start user onboarding ~\cite{sun2023personalization}, where traditional matrix factorization or collaborative filtering fails due to lack of prior data~\cite{ye2023large}. These LLM-driven systems allow users to express their needs directly via natural language, supporting dynamic profile construction and preference elicitation through dialogue. This leads to faster convergence to relevant recommendations and improved first-session satisfaction metrics ~\cite{peng2025surveyllmpoweredagentsrecommender}.

\begin{figure}[H]  % Use [htbp] if you don't want it fixed here
    \centering
    \includegraphics[width=1.0\textwidth]{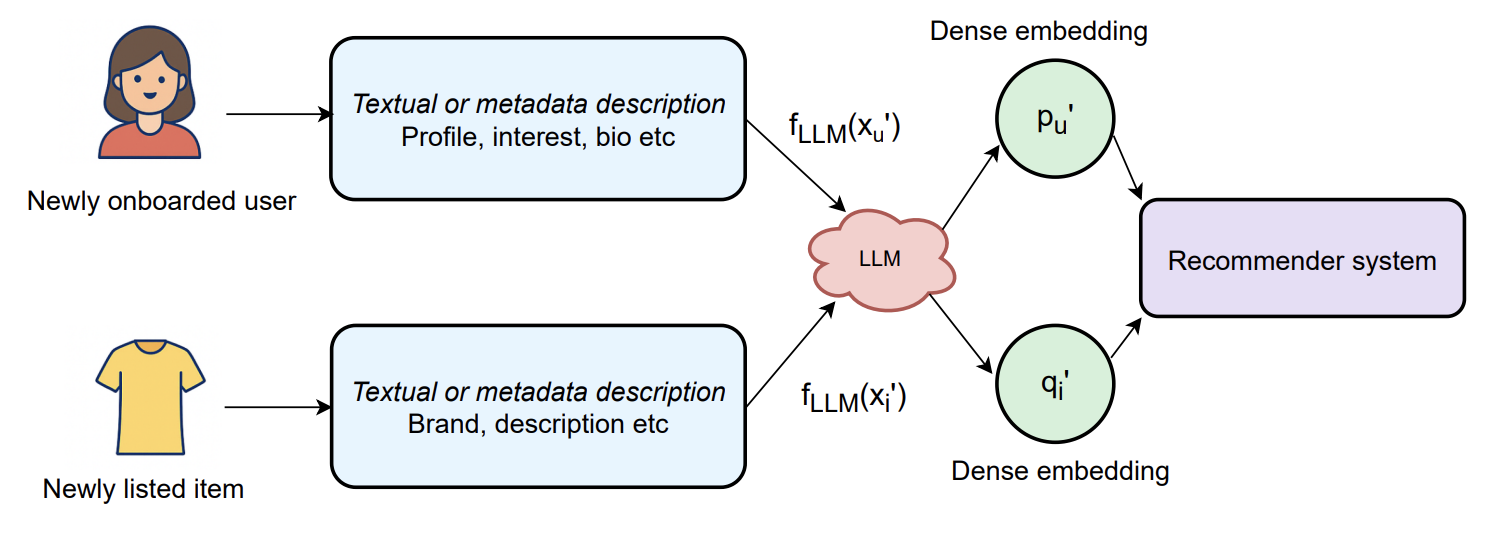}  % Adjust width as needed
    \caption{Illustration of representation bootstrapping solution using LLM to generate dense user and item embeddings from textual or metadata descriptions. Newly onboarded users and newly listed items are mapped to the embedding space via LLM-based encoders, enabling effective integration into the recommender system without historical interaction data.}
    \label{fig:zero_shot_embeddings}
\end{figure}

\textbf{Representation Bootstrapping –}  
LLMs can synthesize dense vector representations for previously unseen users and items directly from textual, profile, or metadata descriptions, offering a lightweight embedding initialization pipeline ~\cite{tennenholtz2024demystifyingembeddingspacesusing}. This capability is particularly valuable in \textit{cold-start} scenarios, where traditional collaborative filtering methods struggle due to the absence of historical interaction data. By leveraging rich semantic priors from pretraining on large corpora, LLMs enable \textbf{zero-shot generalization}, allowing the system to produce meaningful embeddings for novel entities without requiring full model retraining or embedding table updates.~\cite{huber2025embeddingtoprefixparameterefficientpersonalizationpretrained}

For instance, newly listed products on e-commerce platforms or newly onboarded users can be immediately embedded into the same vector space as existing catalog items and users:
\[
\mathbf{q}_{i'} = f_{\text{LLM}}(x_{i'}), \quad \mathbf{p}_{u'} = f_{\text{LLM}}(x_{u'})
\]
where \(x_{i'}\) and \(x_{u'}\) denote available textual or structured metadata (e.g., product titles, descriptions, user bios, or interests). These bootstrapped vectors can be plugged into existing retrieval or ranking models with minimal latency, significantly reducing onboarding time.

Moreover, such LLM-derived embeddings can serve as \textbf{informative priors} during fine-tuning or personalization ~\cite{zhang-etal-2025-personalize}. They can be adapted via techniques like prompt tuning, adapter layers, or shallow re-embedding using user behavior logs, enabling rapid domain alignment while preserving generalization capabilities ~\cite{wang2023coldgpt}. This framework has been shown to outperform static initialization strategies in real-world systems~\cite{tao2024llmseffectiveembeddingmodels}, and aligns with emerging trends toward \textit{retrieval-augmented generation} and \textit{hybrid recommender architectures} that blend semantic understanding with behavioral learning.

\vspace{0.5em}
\textbf{Language-Native Dialogue Systems –}
LLMs as conversational interfaces can dynamically elicit user preferences in natural language through real-time feedback loops, thereby bypassing the need for extensive logged interactions ~\cite{chan2024humanllmbasedvoiceassistant}. Recent systems like DialogRec ~\cite{wang2024dialogrec} extend this with reinforcement learning over user responses.

This paradigm enables \textit{language-native} recommendation systems, where user profiles are not just inferred from passive behavior logs but actively shaped via dialogue history ~\cite{zheng2024unifiedparadigmintegratingrecommendation}. The model continuously updates its belief about the user's intent and preferences through clarifying questions, preference disambiguation, and exploratory queries. Mathematically, the user representation can be modeled as a latent belief state \( \mathbf{b}_u^t \) updated over time:
\[
\mathbf{b}_u^{t+1} = f_{\text{update}}(\mathbf{b}_u^t, a_t, r_t)
\]
where \( a_t \) is the system's action (e.g., a recommendation or query), and \( r_t \) is the user's response interpreted by the LLM.

Such systems can employ few-shot prompting or in-context learning to tailor responses dynamically without task-specific retraining ~\cite{jiang2025longtermmemoryfoundation}. Additionally, integration with RAG modules allows grounding recommendations in real-time inventory or knowledge bases, increasing factual alignment and diversity ~\cite{fan2024surveyragmeetingllms}.

These conversational systems also pave the way for more inclusive recommendation strategies by reducing dependence on historical bias and enabling preference elicitation in underrepresented cohorts ~\cite{kostric2025tailortalkunderstandingimpact}. In high-stakes domains like healthcare, education, or legal support, the ability to engage users in naturalistic conversation supports transparent, explainable, and ethical recommendation pipelines ~\cite{ferdaus2024trustworthyaireviewethical}. As user preferences evolve over time, language-native agents offer a compelling path toward \textbf{continual preference learning} ~\cite{zhang2024coprcontinualhumanpreference} and \textbf{intent-aware retrieval} ~\cite{chen-etal-2023-generate}, making them an ideal front-end for next-generation recommender architectures.

\vspace{0.5em}
\textbf{Prompt-Based Conditioning –} Prompt-based conditioning with LLMs enables flexible zero-shot personalization by translating user-item interactions into natural language templates ~\cite{ramos2025peapodpersonalizedpromptdistillation}. This approach reframes recommendation as a language modeling task, allowing LLMs to directly estimate the likelihood of recommending item \( i \) to user \( u \) using a natural language prompt:
\[
P(i \mid x_u) = \text{LLM}(\texttt{``User context: ''} + x_u + \texttt{`` Recommend: ''} + x_i)
\]
where \( x_u \) and \( x_i \) represent the textual representations (e.g., profile, metadata, reviews) of the user and item, respectively. Unlike traditional scoring models that rely on latent dot-product similarity or supervised learning over explicit interaction labels, this formulation is \textit{instruction-driven} and supports a wide range of open-ended tasks.

By conditioning generation or scoring on structured prompts, this modality-agnostic interface allows for multiple downstream applications: justification of recommendations (e.g., ``why was this item selected?''), prediction of preference explanations (e.g., ``what might the user say about this item?''), and hybrid tasks such as comparative ranking or critique generation. It enables interaction-rich scenarios even in the absence of dense interaction histories, facilitating \textbf{few-shot generalization} in cold-start settings ~\cite{LiuYufang2024AMFf}.

Notably, recent systems like PromptRec~\cite{dai2023promptrec}, PROMO~\cite{jiang2024prompttuningitemcoldstart}, and OpenAI’s plugin-based RecSys prototype~\cite{li2023generative} have operationalized such strategies in both online and offline settings. These systems dynamically compose prompts using structured user/item attributes and contextual cues (e.g., time, location, browsing intent), effectively merging retrieval and generation within a single architecture.

Furthermore, prompt-based approaches are inherently interpretable and modular, offering compatibility with personalization layers, response reranking, or safety filters ~\cite{wang2025automatingpersonalizationpromptoptimization}. Their plug-and-play nature aligns well with \textbf{low-code/zero-code} deployment paradigms, reducing the friction in experimentation and scaling across domains and languages ~\cite{cheng2023plug}. As LLMs continue to improve in understanding long-tail user intents and compositional queries, prompt-based conditioning is emerging as a powerful mechanism for building \textit{controllable, transparent, and context-aware} recommender systems.

\begin{figure}[htb]  % Use [htbp] if you don't want it fixed here
    \centering
    \includegraphics[width=1.0\textwidth]{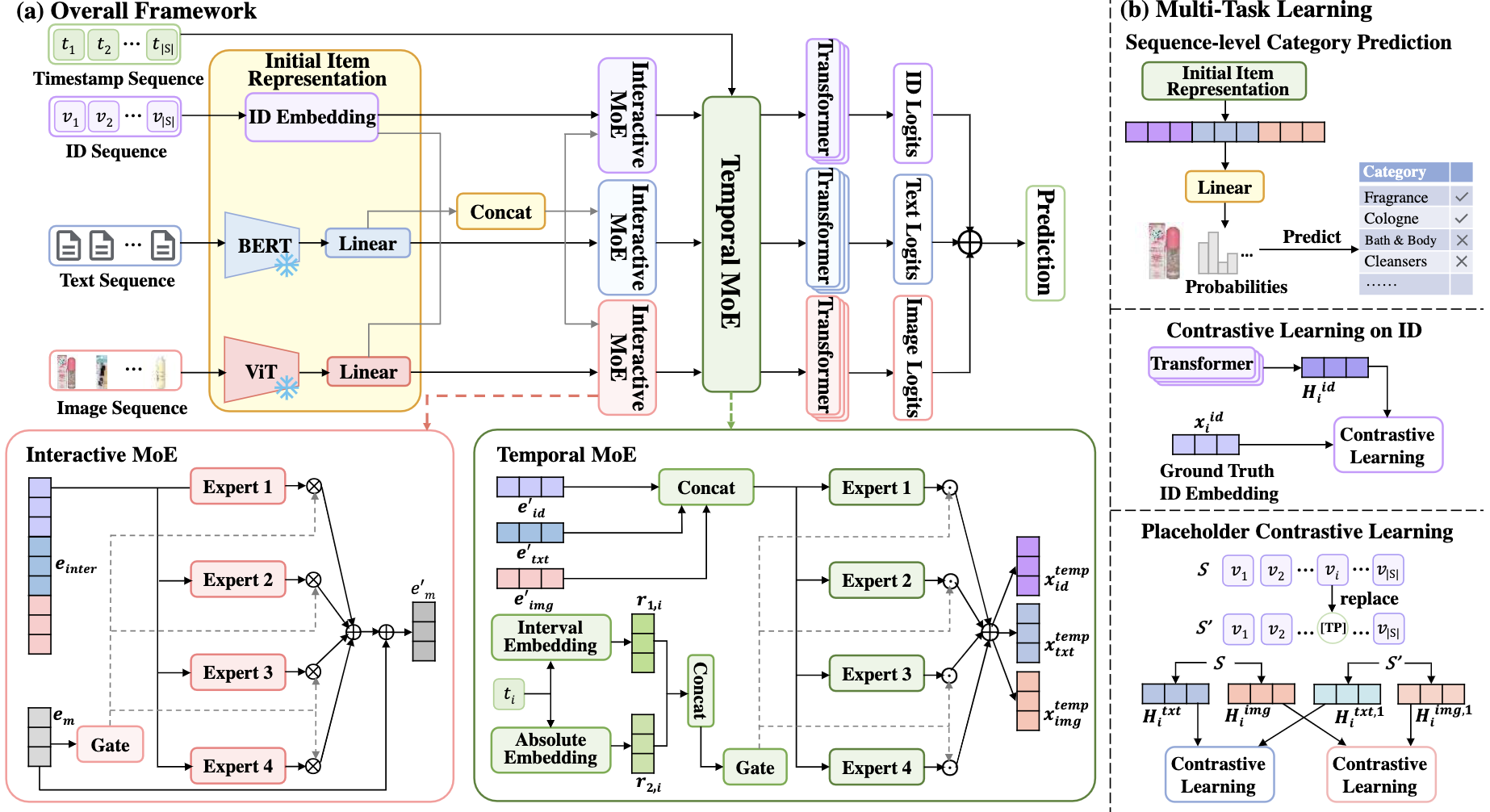}  % Adjust width as needed
    \caption{Overview of the Proposed Multi-Modal Sequence Learning Framework.(a) The model processes timestamp, ID, text, and image sequences using BERT, ViT, and ID embeddings, followed by Interactive and Temporal MoE modules for expert routing. Outputs are fed into Transformers for prediction. (b) Multi-task learning includes category prediction, contrastive learning on ID embeddings, and placeholder-based contrastive learning across modalities ~\citep{zhang2025hierarchicaltimeawaremixtureexperts}.}
    \label{fig:retro-architecture}
\end{figure}

\vspace{0.5em}
\textbf{Multimodal Embedding Synthesis –}When textual information alone is insufficient for high–fidelity personalization, LLMs can be situated at the center of a \emph{multimodal} pipeline that fuses visual, structured, and temporal signals ~\Cref{fig:retro-architecture}).
d behavioral signals into a unified representation ~\cite{gadre2023video}. A generic fusion template is:
\[
\mathbf{q}_i \;=\; f_{\text{text}}\!\bigl(x_i^{\text{text}}\bigr) 
\;+\; f_{\text{img}}\!\bigl(x_i^{\text{img}}\bigr) 
\;+\; f_{\text{meta}}\!\bigl(x_i^{\text{meta}}\bigr),
\]
where $x_i^{\text{text}}$ (e.g., product title, description), $x_i^{\text{img}}$ (primary and auxiliary images), and $x_i^{\text{meta}}$ (price, brand, category, timestamp, CTR priors) are embedded by modality-specific encoders and subsequently \textit{aligned} in a shared latent space.  

State-of-the-art LLMs such as \textsc{Flamingo}~\cite{alayrac2022flamingo} use cross-attention layers to inject image token embeddings into frozen language blocks, while lighter adapters or multi-head cross-modal transformers can accomplish a similar goal with lower inference cost. The fusion function can be further refined with a learned gating vector $\boldsymbol{\gamma}_i$:
\[
\mathbf{q}_i \;=\; \boldsymbol{\gamma}_i^{\top}
\!\begin{bmatrix}
f_{\text{text}}\!\bigl(x_i^{\text{text}}\bigr)\\[2pt]
f_{\text{img}}\!\bigl(x_i^{\text{img}}\bigr)\\[2pt]
f_{\text{meta}}\!\bigl(x_i^{\text{meta}}\bigr)
\end{bmatrix}, 
\qquad \gamma_{i,m} = \frac{\exp(g_m)}{\sum_{m'}\exp(g_{m'})},
\]
where the gating logits $g_m$ are produced by an \emph{attention-over-modalities} layer conditioned on contextual cues (e.g.\ user intent or query).  
This adaptive weighting has two practical advantages: (i) it suppresses noisy modalities when signals are weak (e.g.\ blurry images) and (ii) it accentuates the most informative channel for each request, leading to more stable cold-start performance ~\cite{yu2022modality}. Amazon’s Multimodal RecSys~\cite{hou2023mmrec} employs hierarchical fusion: fine-tuned \textsc{CLIP} encoders supply image vectors, BERT-based encoders supply text, and catalog metadata is featurized via embedding tables. A shallow cross-modal transformer merges these signals and outputs $\mathbf{q}_i$ to a traditional ANN service, yielding a \(\sim\)5–10\% lift in hit-rate for products with \textless{}5 historical interactions ~\cite{chang2021mmrec}. Similar blueprints are emerging in streaming platforms that blend thumbnails, transcripts, and user watch-history to build \textit{rich} item sketches for zero-shot retrieval ~\cite{xu2022unified}. Future work is exploring \emph{co-training} regimes where the LLM simultaneously solves captioning ~\cite{hao2024traininglargelanguagemodels}, visual question-answering, and recommendation; this multitask conditioning regularizes the shared space and mitigates over-fitting to any single modality. Knowledge-distillation and quantization pipelines are being introduced to compress the fused encoders for edge deployment, aligning multimodal personalization with latency and privacy constraints.

\vspace{0.5em}

\begin{figure}[h]
  \centering
  \includegraphics[width=1.0\textwidth]{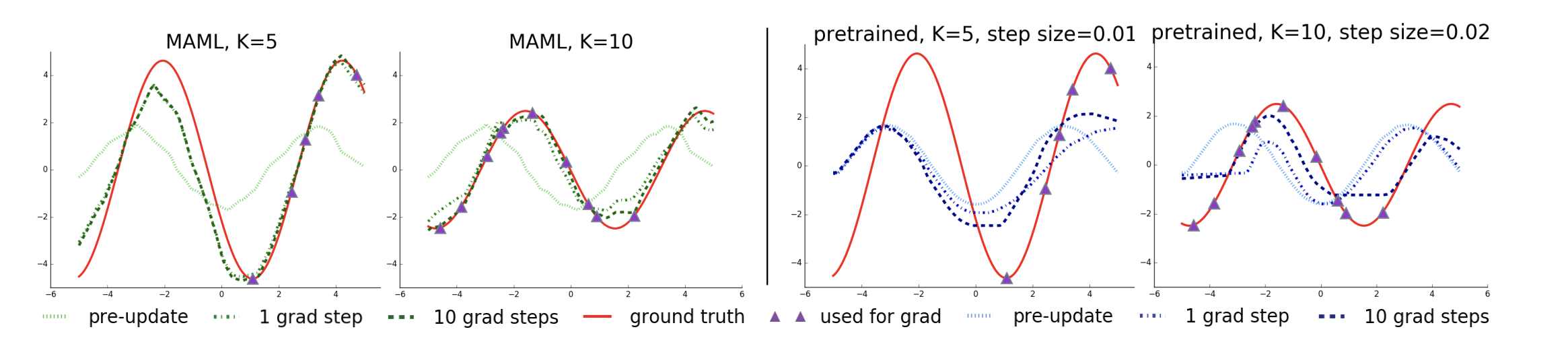}  % Adjust filename as needed
  \caption{The image compares MAML-based few-shot adaptation (left) to standard fine-tuning without MAML (right) for a sine wave regression task. MAML generalizes well, even in regions without data, by capturing the underlying periodic structure. In contrast, the pretrained model struggles to adapt from few examples, failing to extrapolate due to conflicting gradients from pretraining~\citep{finn2017modelagnosticmetalearningfastadaptation}}
  \label{fig:maml}
\end{figure}

\vspace{0.5em}
\textbf{Meta-Learning for Fast Adaptation –} Meta-learning provides a framework for enabling rapid personalization by training models to adapt quickly to new users or items using limited data ~\cite{wang2024limamlpersonalizationdeeprecommender}. When combined with LLMs, meta-learning techniques such as Model-Agnostic Meta-Learning (MAML)~\cite{finn2017maml} allow systems to fine-tune user- or item-specific representations via a small number of gradient updates(see Figure~\ref{fig:maml}. For a new user task \( \mathcal{T}_u \), the model parameters are adapted as follows: \[ \theta^{*} = \theta - \alpha \nabla_{\theta} \mathcal{L}_{\mathcal{T}_u}(f_{\theta}) \], where \( \theta \) are the shared model parameters and \( \alpha \) is the learning rate. This enables the system to personalize quickly without retraining from scratch. Recent works such as \textsc{MeLU}~\cite{lee2019melu} and \textsc{Meta-LLMRec}~\cite{lin2023metallmrec} demonstrate how few-shot fine-tuning on top of LLM-derived embeddings can deliver strong performance in cold-start settings, achieving meaningful generalization with minimal supervision. These methods also support continual learning across user cohorts, enabling fast model updates without overfitting or catastrophic forgetting, and are particularly well-suited for dynamic environments like e-commerce or content streaming platforms ~\cite{hamedi2025federatedcontinuallearningconcepts}.

\vspace{0.5em}
\textbf{Cross-Domain Transfer via Language –} LLMs pretrained on diverse and heterogeneous corpora can facilitate cross-domain recommendation by leveraging their semantic understanding to bridge disparate user-item interaction spaces ~\citep{vajjala2024crossdomainrecommendationmeetslarge, yi2023contrastivegraphprompttuningcrossdomain, Liu_2025,vats2024exploringimpactlargelanguage}. In scenarios where domains $\mathcal{D}_A$ and $\mathcal{D}_B$ have minimal or no overlapping users or items, traditional collaborative filtering fails due to a lack of common signals ~\cite{zhu2025collaborativeretrievallargelanguage}. However, LLMs can align these domains by grounding recommendations in natural language semantics shared across domains. Formally, user preferences learned from $\mathcal{D}_A$ can be transferred to items in $\mathcal{D}_B$ by:

\[
P_{\mathcal{D}_B}(i \mid x_u) \approx \text{LLM}(x_u^{\mathcal{D}_A}, x_i^{\mathcal{D}_B})
\]

This enables zero-shot generalization to new item taxonomies or verticals, using only text, tags, or descriptions ~\cite{sanh2022multitaskpromptedtrainingenables}. For instance, a user who frequently interacts with technical blog posts (in $\mathcal{D}_A$) can be matched with programming tutorials or job listings (in $\mathcal{D}_B$), purely through semantic alignment.

Recent approaches such as PromptRec and LLMRec show that this strategy improves recommendation relevance in cross-modal and multilingual setups. Additionally, systems like CrossAligner~\cite{gritta2022crossaligner} use contrastive fine-tuning to align embeddings across domains, while ~\cite{fu2024comprehensive} demonstrate strong performance on cold-start item transfer using instruction-following LLMs. This line of work opens avenues for applying pretraining-era language representations to cross-domain personalization at scale, eliminating the need for costly alignment mappings or dual-domain retraining.
% \vspace{-4.5em} % Adjust as needed (start small!)
\begin{figure}[H]
  \centering
  \includegraphics[width=1.09\textwidth]{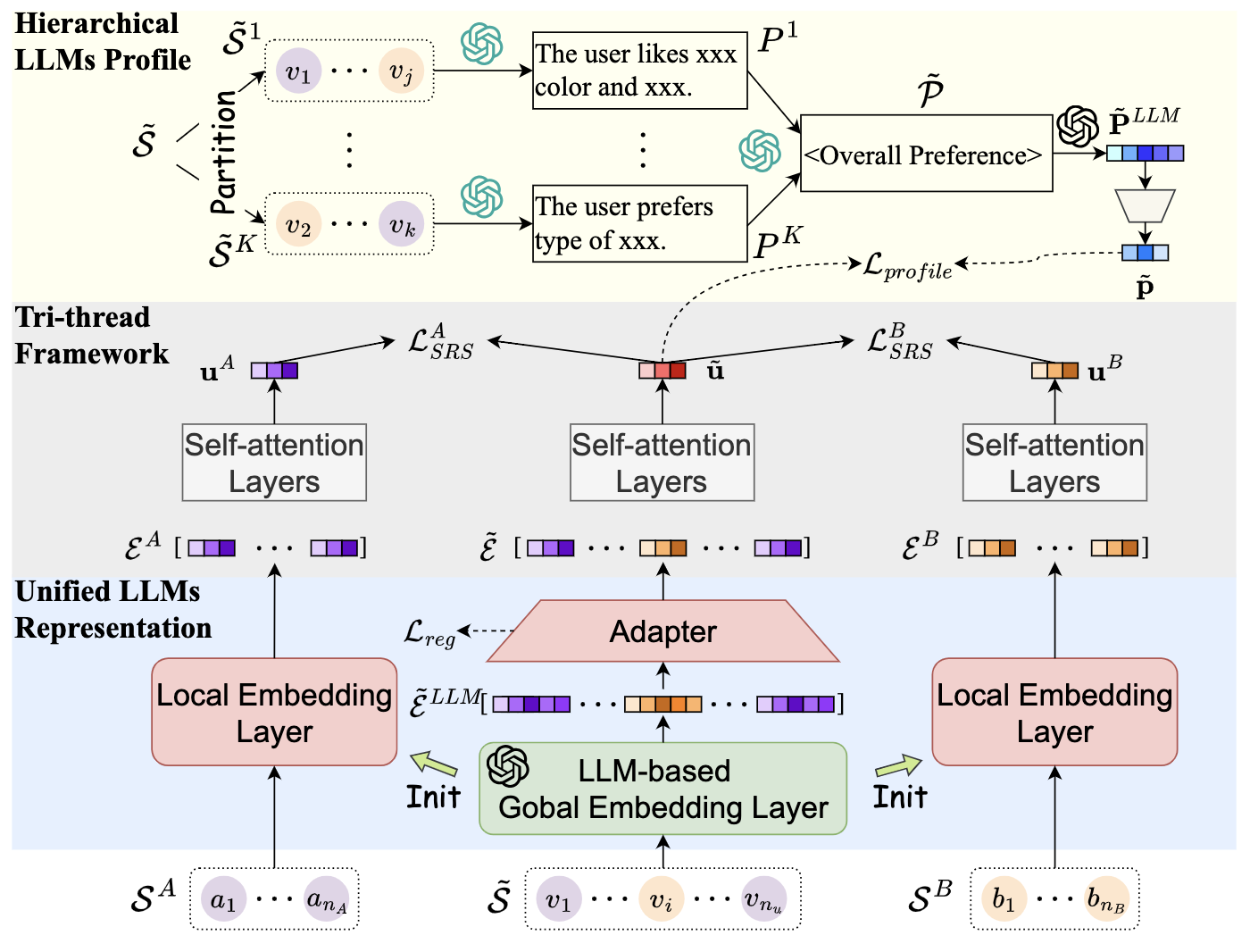}  % Adjust filename as needed
  \caption{Overview of the LLM4CDSR framework for Cross-Domain Sequential Recommendation. The architecture first constructs hierarchical user profiles by partitioning user history and generating summaries using LLMs. A tri-thread framework integrates user behavior across source and target domains through self-attention layers. An adapter connects local embeddings with a global LLM-based representation to enable unified personalization across domains.~\citep{liu2025bridgedomainslargelanguage} }
  \label{fig:temporal-drift}
\end{figure}

\subsubsection{Data Sparsity}

In large-scale recommender systems, the user-item interaction matrix $R \in \mathbb{R}^{|U| \times |I|}$ is predominantly sparse, with observed interactions constituting less than 0.1\% of the total possible interactions. The sparsity ratio is quantified as:
\[
\text{Sparsity} = 1 - \frac{\|\mathcal{R}_{\text{obs}}\|_0}{|U| \cdot |I|}
\]
where $\mathcal{R}_{\text{obs}}$ denotes the set of observed interactions. This high level of sparsity hampers the performance of collaborative filtering techniques, which rely on sufficient interaction data to model user preferences effectively.

Conventional methods to mitigate data sparsity include matrix factorization techniques such as Singular Value Decomposition (SVD) \cite{1102314} and Non-negative Matrix Factorization (NMF) \cite{lee1999learning}, as well as hybrid models that incorporate side information \cite{melville2002content,adomavicius2005toward}. While these approaches can alleviate sparsity to some extent, they often struggle with scalability and may not capture complex user-item relationships, especially in scenarios with extremely sparse data \cite{vahidy2021resolving}. Let's discuss how LLMs can solve the data sparsity issue in recommender systems.

\vspace{0.5em}
\textbf{Text-Driven Generalization –} Pretrained LLMs encode extensive linguistic and factual knowledge from large-scale corpora, enabling semantic reasoning over textual representations of users and items. Rather than relying solely on historical interaction signals (e.g., clicks, views, ratings), these models infer user-item relevance by computing semantic alignment between natural language metadata, leveraging contextual understanding rather than frequency-based heuristics.

This approach is particularly effective in \emph{sparse settings}, where traditional collaborative filtering methods struggle due to the lack of sufficient interaction history. Given item metadata $x_i$ (e.g., title, abstract, tags) and user profile $x_u$ (e.g., occupation, interests, or past behaviors), the model constructs a prompt and estimates a relevance score as:

\[
\hat{r}_{ui} = \mathrm{LLM}(\text{``User profile: ''} + x_u + \text{`` Item: ''} + x_i),
\]

By framing the recommendation problem as a zero-shot or few-shot prediction task, this formulation enables generalization to previously unseen user-item pairs $\langle u, i \rangle$ that are entirely missing from training data. The LLM infers relevance based on semantic similarity, textual context, and its pretrained knowledge.

For example, ~\cite{zhang2021b}and \citet{chen2023prompt} investigate prompt-based LLM scoring for sparse recommendation proposes , a generative framework that predicts user preferences using descriptive prompts. Similarly, \textsc{GENREC}~\citep{cui2023genrecs} extends this idea to fully generative recommender systems that synthesize user-item relevance in the absence of interaction logs.

\vspace{0.5em}
\textbf{Semantic Matching via Embedding Generation -} LLMs can be employed to generate high-dimensional, semantically meaningful embeddings from textual inputs, facilitating effective user-item matching in the absence of interaction data. In this paradigm, LLMs act as encoders that map user profiles and item metadata into a shared latent space, enabling recommendations through embedding similarity. For previously unseen users or items, the relevance score is computed as:

\[
\hat{p}_u = f_u(x_u), \quad \hat{q}_i = f_i(x_i), \quad \hat{r}_{ui} = \hat{p}_u^\top \hat{q}_i,
\]

\noindent where $f_u$ and $f_i$ are LLM-based encoders applied to the user profile $x_u$ and item description $x_i$, respectively. These encoders can leverage instruction tuning, adapter modules, or lightweight prompt-based techniques to incorporate domain-specific context during representation learning.

This semantic matching framework has demonstrated strong performance in cold-start and zero-shot settings ~\citep{wang2023llm4coldstart} uses LLMs to encode textual features for new users and items, enabling retrieval via embedding similarity~\citep{cui2024promptbasedknowledgegraphfoundation}, this further incorporates structured knowledge graph information via prompt tuning to improve recommendation accuracy.\\

\subsubsection{Noisy Implicit Feedback}

Modern recommender systems predominantly rely on implicit feedback due to the scarcity of explicit labels. Let $o_{ui} \in \mathcal{O}$ denote an observed implicit interaction between user $u$ and item $i$ (e.g., click, impression, dwell), and $r_{ui}^{*} \in \{0, 1\}$ be the unobserved ground-truth relevance. In theory, one might assume:
\[
\mathbb{P}(o_{ui} \mid r_{ui}^{*}) = \delta(r_{ui}^{*}),
\]
where $\delta$ denotes a deterministic mapping. However, in practice, this assumption fails due to multiple sources of stochasticity and bias in user behavior. The actual observed feedback often satisfies:
\[
\mathbb{P}(o_{ui} = 1 \mid r_{ui}^{*} = 1) < 1, \quad
\mathbb{P}(o_{ui} = 1 \mid r_{ui}^{*} = 0) > 0.
\]

These deviations arise from position bias, presentation bias, and selection bias, which skew click-through data toward superficial engagement signals ~\cite{li2024llmsasjudgescomprehensivesurveyllmbased}. This noise propagates into learning algorithms, especially when optimizing binary cross-entropy loss:
\[
\mathcal{L}_{\text{BCE}} =
- \sum_{(u,i) \in \mathcal{O}} \left[
o_{ui} \log \hat{r}_{ui} +
(1 - o_{ui}) \log(1 - \hat{r}_{ui})
\right],
\]
where $\hat{r}_{ui}$ denotes the predicted relevance. Correcting such noisy supervision without access to counterfactual data remains a central challenge.

LLMs pretrained on large-scale human-authored corpora encode rich priors about user intent, preference, and satisfaction—grounded in linguistic patterns and world knowledge ~\cite{bodonhelyi2024userintentrecognitionsatisfaction}. These priors can be leveraged to mitigate the impact of noisy implicit feedback by smoothing binary interaction labels (e.g., clicks or skips) into soft, continuous scores that better reflect underlying user preferences. Specifically, given item content $x_i$ and user context $c_u$ (e.g., prior interactions, demographics, or interest descriptors), an LLM can estimate a denoised relevance score as:
\[
\tilde{r}_{ui} = \text{sigmoid}(\text{LLM}(x_i, c_u)),
\]
where the LLM is prompted to infer how well the item matches the user’s intent. The output $\tilde{r}_{ui} \in (0,1)$ can be interpreted as a soft label and used to replace or augment the original binary signal $o_{ui}$ in the loss function during training. This reformulation introduces robustness by attenuating spurious clicks or missed positives, effectively serving as a learned reweighting scheme.

This denoising strategy has been applied in systems, where LLMs provide feedback-aware relevance estimates for training ranking models. Other works like \textsc{PromptTuning4Rec}~\citep{wang2023prompttuning} and \textsc{LightRec}~\citep{zhang2023lightrec} explore LLM-driven soft label generation as a mechanism to denoise interaction matrices under extreme sparsity. The approach aligns with broader trends in label refinement~\citep{lee2013pseudo, hu2016label} and weak supervision for recommender systems. The next sub-section talks about how LLMs can help in alleviating noisy implicit feedback in recommender systems.

\vspace{0.5em}
\textbf{Feedback Interpretation with Contextual Prompts –}  
LLMs can be used to interpret user interactions by incorporating behavioral context into prompts—such as item position, dwell time, or scroll depth—to assess the reliability or intent behind an observed action. For instance, given a prompt like:

\begin{quote}
\texttt{User clicked item $x_i$ at position 8 after scrolling for 2 seconds. Was this click intentional?}
\end{quote}

\noindent the LLM returns a binary or scalar prediction $\hat{y}_{ui} \in \{0, 1\}$ (or a confidence score), which can then be used for downstream tasks such as sample filtering, importance weighting, or bias correction ~\cite{kruspe2024detectingunanticipatedbiaslarge}. This method enables context-aware re-labeling of interactions and helps distinguish between genuine interest and behavior driven by interface bias (e.g., accidental clicks on top-ranked items or skip-throughs). It builds on the idea that implicit feedback is inherently noisy and requires situational reasoning to infer its semantic relevance ~\cite{yi2025ironydetectionreasoningunderstanding}.

Recent implementations of this strategy where user behavior traces are used to condition prompts that refine training labels; behavior-aware reasoning in \textsc{SAM}~\citep{jin2023sam}, which incorporates dwell and revisit signals into expert routing; and click attribution models that interpret interactions using rich context.Generative LLMs can simulate unseen feedback under counterfactual scenarios. For example:
\begin{quote}
\texttt{If item $x_i$ were shown first instead of last, would the user have clicked?}
\end{quote}
This models counterfactual relevance:
\[
\hat{r}_{ui}^{\text{cf}} = \text{LLM}(\texttt{Counterfactual Prompt}),
\]
providing a proxy for $\mathbb{P}(r_{ui}^{*} \mid \text{do}(x_i^{\text{pos}} = 1))$. LLMs can classify behavior sequences using natural instructions using prompts.  The resulting label $\hat{y}_{ui}$ can be used for fine-tuning or distillation. This approach has been adopted in multi-objective recsys systems like InstRec~\cite{wang2023instrec} and human-feedback alignment setups. LLMs can simulate realistic interaction sequences for pretraining purposes. For example:
\begin{quote}
\texttt{User browsed 5 items. Skipped $x_i$ after reading the summary. Clicked $x_j$ after 3 seconds.}
\end{quote}
Such pseudo-sessions enable robust pretraining across noisy behavioral regimes. This has been used in SAM~\cite{jin2023sam}, LLM4RecSim~\cite{wang2024llm4recsim}, and language-driven session modeling~\cite{xu2023llmr}. LLMs can generate explanations to help identify noisy or biased samples.  These explanations can be post-processed into logic rules for filtering or label smoothing. Explanation-enhanced supervision has been studied in GPT4Rec~\cite{zhou2023gpt4rec}, dialog-based models, and instruction-aligned filtering setups~\cite{gao2023llm4kg}.

\begin{figure}[h]
  \centering
  \includegraphics[width=0.7\textwidth]{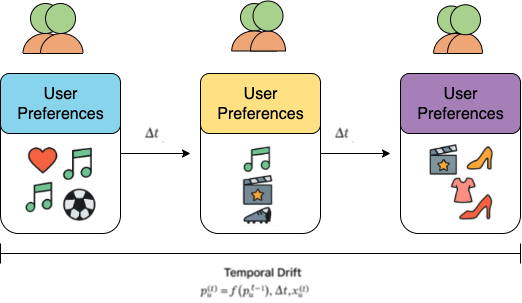}  % Adjust filename as needed
  \caption{Illustration of Temporal Drift in User Preferences. The figure shows how user preferences evolve over time—from the past to the present and into the future. As the user’s interests shift (e.g., from sports to music to fashion), the recommendation system must model this drift by adapting the user's preference representation \( p_u^{(t)} \) as a function of past preferences \( p_u^{t-1} \), elapsed time \( \Delta t \), and contextual features \( x_u^{(t)} \). Capturing this dynamic behavior is crucial for maintaining long-term recommendation relevance.}
  \label{fig:temporal-drift}
\end{figure}

\subsubsection{Temporal Drift}

User preferences are inherently dynamic. Static user embeddings, often learned from aggregate histories, fail to capture shifts in taste caused by evolving life contexts, seasonality, or trending content. A time-aware user representation can be modeled as:
\[
\mathbf{p}_u^{(t)} = f(\mathbf{p}_u^{(t-1)}, \Delta t, \mathbf{x}_u^{(t)})
\]
where $\mathbf{p}_u^{(t)}$ is the user embedding at time $t$, $\Delta t$ denotes the time elapsed since the last interaction, and $\mathbf{x}_u^{(t)}$ includes recent interaction context. Here is how LLMs can help in addressing temporal drift.

\vspace{0.5em}
\textbf{Temporal Adaptation with LLMs}
Recent advances in large language models (LLMs) enable dynamic user modeling via textual prompts that encode temporal context~\cite{zhao2025videoexpertaugmentedllmtemporalsensitive}. Rather than retraining models to reflect evolving preferences, LLMs support \textit{on-the-fly personalization} by accepting expressive, context-aware prompts~\cite{kumar2025leveragingknowledgegraphsllms}. These prompts are designed to capture temporal signals such as recency, session history, preference shifts, and explanatory cues. 

For example, recent user behavior can be embedded into prompts that guide the LLM in generating updated user representations. This mechanism has been explored in PromptRec, LMRec, and genre-adaptive recommendation models~\cite{chai2022genrelongtail}. Similarly, attention over long-term and short-term preferences can be introduced by structuring prompts that separately convey enduring interests and recent actions. LLMs can also be prompted to produce interpretable justifications based on recency, making them suitable for explainable recommendation settings.

In session-aware contexts, prompts constructed from short-term interaction sequences (e.g., clicks, watches, searches) enable the model to simulate user intent in real time and generate slates accordingly. Additionally, instruction-tuned LLMs can perform few-shot drift adaptation by observing behavioral transitions across users and generalizing to new preference shifts without explicit fine-tuning. Noisy or fragmented histories can also be abstracted into stable, high-level summaries through summarization-style prompts.

These strategies are systematically summarized in Table~\ref{tab:prompt_temporal}, which outlines the types of prompts used, their intended purposes, and representative systems that have adopted them.

\begin{table}[ht!]
\centering
\scriptsize
\begin{tabular}{p{6.2cm}|p{4.5cm}}
  \toprule
  \textbf{Prompt Example} & \textbf{Purpose} \\
  \midrule
  \texttt{User watched: [Horror A], [Horror B], [Horror C]. Recommend based on their recent interests.} 
  & Refresh user representation based on recent behavior \\
  \midrule
  \texttt{Long-term favorites: Sci-Fi, Documentaries. Recent watches: Romantic comedies.} 
  & Temporal attention over long-term and short-term interests \\
  \midrule
  \texttt{Because you recently watched romantic comedies, we're recommending this title.} 
  & Generate interpretable, recency-aware explanations \\
  \midrule
  \texttt{Session: clicked [Horror A], watched [Horror B], searched "ghost movies"} 
  & Model short-term session context for slate generation \\
  \midrule
  \texttt{User 1: Horror → Comedy.} \newline \texttt{User 2: Thriller → Animation. Target: Romance → ?} 
  & Enable few-shot drift generalization across users \\
  \midrule
  \texttt{Summarize recent user behavior and extract long-term interests.} 
  & Smooth noisy sequences into high-level intent \\
  \bottomrule
\end{tabular}

\vspace{0.5em}
\caption{Prompt-Based Techniques for Temporal Adaptation in LLM-Based Recommenders}
\label{tab:prompt_temporal}
\end{table}

\vspace{0.5em}
\textbf{Structural and Hybrid LLM Integration}
LLMs can be integrated into recommendation pipelines through structural means that support temporal adaptation ~\cite{luo2024integratinglargelanguagemodels}. One such approach involves using LLMs to parameterize updates to the user embedding. For example, recent behavioral context can be encoded by the LLM and merged with historical representations via a learned transformation:

\[
\mathbf{p}_u^{(t)} = \text{MLP}(\mathbf{p}_u^{(t-1)} \,\|\, \text{LLM}(\mathbf{x}_u^{(t)}))
\]

This enables fine-grained updates to latent factors without explicit prompting during inference. In another variant, LLMs augment sequential recommendation models by enriching item representations or positional embeddings within transformer-based architectures ~\cite{Zivic_2024}. Here, interaction sequences are processed as:

\[
\mathbf{h}_t = \text{Transformer}([\mathbf{x}_1, \ldots, \texttt{LLM}(\mathbf{x}_t)])
\]

which allows the model to incorporate both learned behavior encodings and contextual language insights. Retrieval-augmented approaches also benefit from LLM integration. By conditioning retrieval queries on temporally relevant cues, LLMs dynamically populate the candidate pool based on evolving user contexts ~\cite{han2025rethinkingllmbasedrecommendationsquery}. LLMs also act as soft-labeling supervisors in distillation frameworks ~\cite{sutanto2024llmdistillationefficientfewshot}. For instance, in short or noisy sessions, they generate target summaries that guide smaller models to learn smoothed, temporally aligned representations. 

\subsubsection{Multimodal Data Integration}

Modern recommendation platforms such as Amazon, Pinterest, and Spotify leverage heterogeneous and semantically rich multimodal signals to enhance personalization. These signals include textual descriptions ($x^{\text{text}}$), visual content ($x^{\text{img}}$), structured metadata ($x^{\text{meta}}$), and user interaction histories or behavioral traces ($x^{\text{behav}}$). To produce a unified representation for an item $i$, a common formulation involves modality-specific encoders whose outputs are aggregated to yield the final item embedding:

\[
\mathbf{q}_i = f(x_i) = f_{\text{text}}(x^{\text{text}}_i) + f_{\text{img}}(x^{\text{img}}_i) + f_{\text{meta}}(x^{\text{meta}}_i) + f_{\text{behav}}(x^{\text{behav}}_i),
\]

\noindent where each $f_{\cdot}$ is a modality-specific encoder (e.g., a transformer for text, a CNN for images, a feedforward network for metadata), and $x_i = \{x^{\text{text}}_i, x^{\text{img}}_i, x^{\text{meta}}_i, x^{\text{behav}}_i\}$ denotes the full multimodal feature set associated with item $i$.

However, this fusion process is complicated by challenges such as modality imbalance—where some inputs (e.g., reviews) are informative while others (e.g., short captions) are weak—feature misalignment due to semantic or temporal inconsistencies across modalities, and missing or noisy inputs, particularly in cold-start scenarios where behavioral signals may be absent or unreliable.

These issues limit the effectiveness of conventional fusion methods such as early concatenation or late summation, particularly in sparse or cold-start regimes. The next subsections disccuss the approaches using LLM to solve the multimodal data integration issues.

\vspace{0.5em}
\textbf{Modal-Aware Weighting through Attention –}  
Transformer-based LLMs naturally address heterogeneity via \emph{modal-aware attention mechanisms}, dynamically re-weighting modalities during representation learning \citep{xu2023llmr,jin2023sam,ma2024xrec}.  
Self-attention computes token-level dependencies both across and within modalities. When structured attributes (e.g., brand, category) are incomplete, the model shifts focus toward unstructured signals such as reviews or captions; conversely, when textual cues are weak, more attention mass is allocated to reliable metadata or visual embeddings. This \emph{soft modality selection} is realized implicitly, without explicit gating or masking heuristics ~\cite{Cai_2025}.

\begin{figure}[h]
  \centering
  \includegraphics[width=1.0\textwidth]{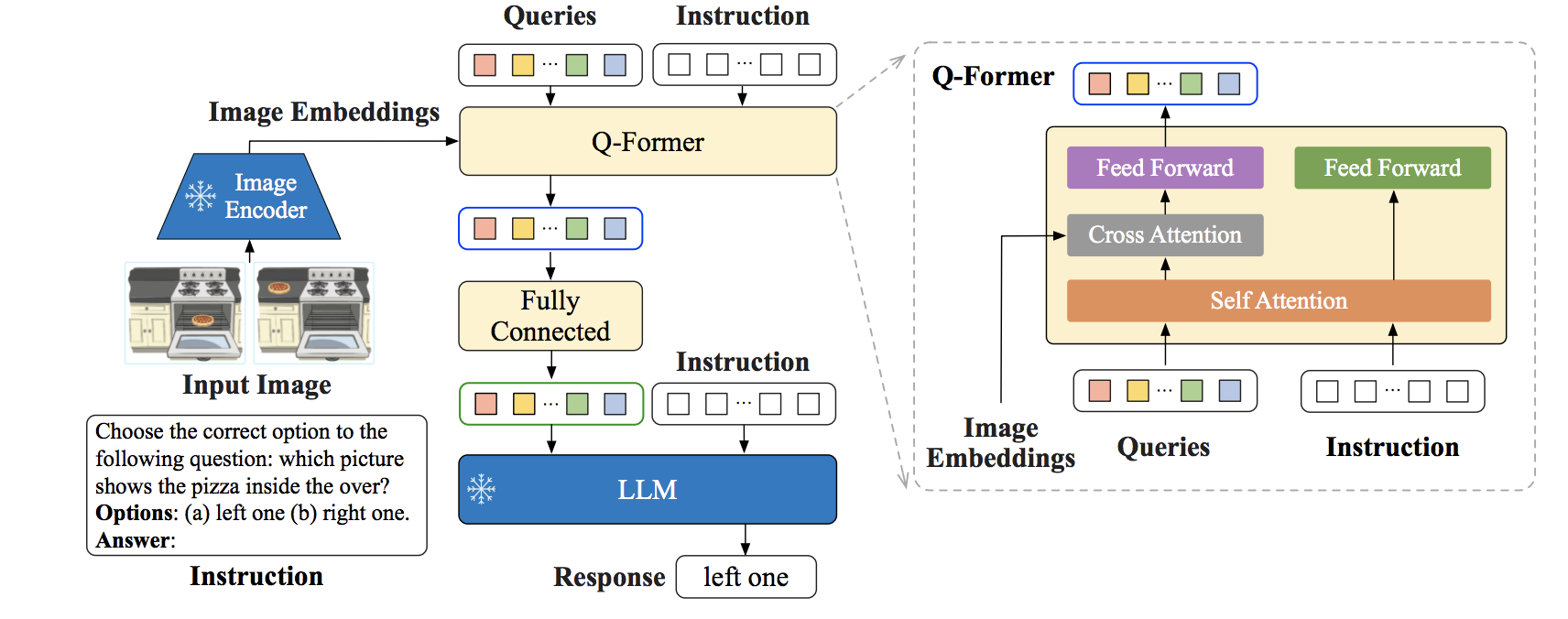}  % Adjust filename as needed
  \caption{An instruction-tuned LLM framework that unifies visual and textual inputs into a shared semantic space. The image encoder extracts visual embeddings, which are fused with textual instructions via a Q-Former module using self-attention and cross-attention. The combined representation is processed by the LLM to generate a response, enabling cross-modal reasoning without separate modality-specific towers.~\citep{dai2023instructblipgeneralpurposevisionlanguagemodels}}
  \label{fig:instruction-llm}
\end{figure}

\textbf{Unified Representation through Instruction-Tuned LLMs –}  
Given a structured prompt $\texttt{Prompt}_i$ that serializes all modality features (e.g., title, description, image caption, metadata, and recent user feedback), an instruction-tuned LLM(see Figure~\ref{fig:instruction-llm} can embed this multimodal input using a shared semantic space. The dense item representation is computed as:

\[
\mathbf{q}_i = g_{\text{pool}}\!\bigl(\mathrm{LLM}(\texttt{Prompt}_i)\bigr),
\]

\noindent where $g_{\text{pool}}$ typically refers to mean pooling over token embeddings or \textsc{[CLS]}-token extraction. Instruction tuning aligns the LLM's behavior with domain-specific objectives, ensuring that the model respects segment-level distinctions while enabling cross-modal reasoning within a unified encoder. This avoids the need for separate modality-specific towers or handcrafted fusion heuristics.

This paradigm supports scalability, simplifies architecture, and enables zero-shot or few-shot personalization. It has been adopted in \textsc{UniModalRec}~\citep{lu2022unified} for retail recommendation,  and \textsc{UnifiedIO}~\citep{lu2023unifiedio2scalingautoregressive} for vision-language tasks for streaming media. Moreover, \textsc{InstructRecLM} demonstrates that parameter-efficient instruction-tuned LLMs can achieve competitive performance even in low-resource or cold-start regimes.

\vspace{0.5em}
\textbf{Cross-Modal Alignment via Pretraining –}  
Large-scale pretraining on multimodal corpora (e.g., alt-text–image pairs, metadata–review co-occurrences) enables LLMs to implicitly align semantically related content across modalities ~\cite{sapkota2025multimodallargelanguagemodels}. These pretrained models induce a joint latent space where concepts are consistently grounded across different input forms ~\cite{zhang2025unifiedmultimodalunderstandinggeneration}. At inference, when certain modalities are unavailable, the model can still generate meaningful representations by conditioning on the observed subset $x_i^{\text{obs}} \subset x_i$:

\[
\mathbf{q}_i = g_{\text{pool}}\!\bigl(\mathrm{LLM}(\pi(x_i^{\text{obs}}))\bigr),
\]

\noindent where $\pi(\cdot)$ formats available features into a cross-modal prompt. The model fills gaps by relying on shared linguistic anchors (e.g., a product’s visual properties inferred from its textual reviews or structured tags).

This alignment strategy originated in vision-language models like \textsc{CLIP} ~\cite{radford2021learning} and \textsc{ALIGN} ~\cite{jia2021scaling}, and has been adapted for recommendation in \textsc{PaLI}~\citep{chen2023pali}, \textsc{CoCa-2Rec}~\citep{wang2024cocarec}, and \textsc{OmniFM}~\citep{zhou2025omnifm}, where the model generalizes across image-text-review-metadata domains with minimal supervision.

\vspace{0.5em}
\textbf{Multimodal Imputation via Generative Inference –}  
In sparse environments, generative LLMs can hallucinate plausible approximations for missing modalities based on contextual conditioning. When inputs such as user reviews or image captions are missing, the model autoregressively generates substitute content from the conditional distribution:

\[
p_\theta\!\bigl(x_i^{\text{miss}} \mid x_i^{\text{obs}}\bigr) = \prod_{t=1}^{T} p_\theta(w_t \mid w_{<t}, x_i^{\text{obs}}),
\]

\noindent where $x_i^{\text{obs}}$ includes observed fields such as title and metadata, and $(w_1, \dots, w_T)$ are tokens representing the imputed modality (e.g., a visual scene description). This generated text can either be appended to the original prompt or passed through modality-specific encoders, enhancing downstream coverage ~\cite{jung2025learninggeneralizablepromptclip}. This approach enables robust pretraining and fine-tuning in low-resource or noisy-input conditions. 

\vspace{0.5em}
\textbf{Semantic Fusion via Natural-Language Templates –}  
To unify modality inputs without custom encoder branches, each modality is embedded into structured natural language using templated segments. For instance:

\begin{quote}
\texttt{[TITLE] Red running shoes. [IMAGE] Person running on track. [META] Size 9; lightweight. [REVIEW] Very comfortable.}
\end{quote}

\noindent This string is fed into a standard LLM, which processes the full multimodal context as a single token sequence. Reserved tags (e.g., \texttt{[META]}, \texttt{[REVIEW]}) act as soft positional anchors, helping self-attention layers model dependencies both within and across modalities.

Unlike traditional late-fusion models that aggregate fixed embeddings, this strategy offers a fully end-to-end formulation. It is leveraged in \textsc{CM3}~\citep{aghajanyan2022cm3causalmaskedmultimodal}, expanded in \textsc{PromptFusionRec}~\citep{dai2024muapmultistepadaptiveprompt}, and coupled with retrieval-augmented generation in \textsc{UnifiedIO}, facilitating scalable multimodal reasoning while preserving alignment across modality-specific semantic structures.

\subsection{Modeling and Algorithmic Challenges}

Recommender systems today need to deliver a highly personalized user experience with industrial-scale operations. However, the modeling and algorithmic components of these systems are plagued with natural limitations in performance, generalizability, and fairness. In this section, five key bottlenecks for modeling are explored, personalization vs. generalization, scalability of deep models, long-tail modeling, contextual and sequential understanding, and bias amplification, and industry and academia's advancements toward overcoming these bottlenecks are discussed.
\subsubsection{Personalization vs. Generalization}
Recommender systems face an inherent trade-off between personalization and generalization. Personalization aims to tailor recommendations uniquely to each user based on past interactions, whereas generalization ensures the model performs robustly for new users, unseen items, or domain shifts. Overpersonalization can lead to overfitting, where the model memorizes idiosyncratic preferences without learning generalizable patterns. This is especially detrimental in dynamic contexts such as career transitions or shifting media consumption.

Mathematically, overfitting arises when the training objective focuses solely on minimizing personalized reconstruction error:
\begin{equation}
\mathcal{L}_{\text{overfit}} = \sum_{u \in \mathcal{U}} \sum_{i \in \mathcal{I}_u} \left( \hat{r}_{ui} - r_{ui} \right)^2,
\label{eq:overfit_loss}
\end{equation}
where $\hat{r}_{ui}$ is the predicted interaction (e.g., rating or click) and $r_{ui}$ the observed label. Without external context or structural priors, the model fails to generalize beyond $\mathcal{I}_u$—the items historically interacted with by user $u$.

LLMs mitigate the personalization-generalization trade-off through a unified prompt-based interface that integrates multiple strategies into a cohesive framework. \textbf{Instruction-Tuned Generalization} exploits the global instruction-following capabilities of models such as \textsc{FLAN-T5}~\cite{chung2022scaling}, \textsc{InstructGPT}~\cite{ouyang2022instructgpt}, and \textsc{T0}~\cite{sanh2022multitaskpromptedtrainingenables}, enabling dynamic adaptation to new contexts without retraining. \textbf{Prompt-Tuned Personalization} enhances recommendations by directly embedding structured user feedback within prompts, thereby removing the necessity for continuous fine-tuning; approaches like \textsc{PROMO}~\cite{jiang2024prompttuningitemcoldstart} and \textsc{PEPLER}~\cite{li2022pepler} successfully leverage explicit user preference signals. Additionally, \textbf{Behavioral Diversity via Generative Modeling} generates diverse recommendation candidates using generative prompts, fostering exploration and preventing recommendation collapse—a strategy notably implemented in Spotify's mood-based playlists and e-commerce diversification strategies employed by platforms such as JD and Alibaba. Finally, \textbf{Multitask Prompting} simultaneously encodes multiple recommendation objectives, aligning LLM outputs with multitask learning paradigms, thereby enhancing interpretability and robustness, especially in sparse data environments. This approach is exemplified by systems like RecDSS~\cite{houlsby2019parameter} and instruction-based recommendation tutors~\cite{wang2023instrec}. Table~\ref{tab:rec_prompts} summarizes canonical prompt templates associated with these strategies.
\begin{table}[ht!]
\centering
\scriptsize
\begin{tabular}{p{4.2cm}|p{6.0cm}}
  \toprule
  \textbf{Prompting Strategy} & \textbf{Example Prompt} \\
  \midrule

  Instruction-Tuned Generalization 
  & \texttt{User recently changed career to data science. Suggest jobs based on new interests.} \\
  \midrule

  Prompt-Tuned Personalization 
  & \texttt{User likes: \textless{}liked\_items\textgreater{}. Recommend: \textless{}candidate\_item\textgreater{}} \\
  \midrule

  Behavioral Diversity via Generative Modeling 
  & \texttt{Generate \textless{}n\textgreater{} diverse recommendations for: \textless{}user\_profile\textgreater{}} \\
  \midrule

  Multitask Prompting 
  & \texttt{Recommend an item, explain why it is relevant, and predict how likely the user is to engage.} \\
  \bottomrule
\end{tabular}
\vspace{0.5em}
\caption{ Canonical Prompt Templates for LLM-Based Recommendation Strategies. Angle-bracketed tokens represent runtime placeholders.}
\label{tab:rec_prompts}
\end{table}

Collectively, through \textbf{semantic reasoning}, \textbf{prompt-based control}, and \textbf{generative flexibility}, LLM-based recommenders substantially enhance their ability to generalize across sparse and cold-start scenarios, while maintaining personalized relevance.

% \begin{figure}[ht]
%   \centering
%   \includegraphics[width=0.6\linewidth]{pers-gen.png}
%   \caption{Prompt-based strategies for balancing personalization and generalization using LLMs.}
%   \label{fig:pers-gen}
% \end{figure}

\subsubsection{Scalability of Deep Models}

Deep neural networks (DNNs), transformer-based architectures, and large language models (LLMs) are highly expressive and capable of modeling complex user–item interactions. However, their computational cost makes them impractical for real-time recommendation at industrial scale. The inference latency $\tau$ can be modeled as:
\begin{equation}
\tau = \sum_{l=1}^{L} C_l \cdot d_l,
\label{eq:latency}
\end{equation}
where $L$ is the number of layers, $C_l$ is the compute cost (e.g., FLOPs) of layer $l$, and $d_l$ is the deployment-related delay (e.g., due to memory I/O or parallelism). As model complexity increases, $\tau$ can exceed production constraints.

Two-stage ranking pipelines are widely adopted to alleviate this problem~\cite{perez2021}. First, a lightweight candidate generator $g_{\text{retrieval}}$ retrieves a small set $\mathcal{C}(u)$ from the item pool, followed by a heavier ranking model $f_{\text{rank}}$:
\begin{equation}
\text{Top-}k = \arg\max_{i \in \mathcal{C}(u)} f_{\text{rank}}(u, i).
\label{eq:two_stage}
\end{equation}
While this reduces overall latency, the challenge persists when incorporating deep user or item encoders. Let's see how scalability challenge can be addressed using LLMs.

\textbf{LLM-Based Distillation –}  
Model distillation compresses a large teacher into a smaller student model, reducing latency while preserving performance. For LLMs, distillation does more than parameter compression — it transfers semantic generalization, instruction-following behavior, and multimodal reasoning. This is formalized as:
\[
\mathcal{L}_{\text{distill}} = \sum_{x \in \mathcal{D}} \text{KL}\left(p_{\text{LLM}}(y \mid x) \parallel p_{\text{student}}(y \mid x)\right),
\]
where $p_{\text{LLM}}$ denotes the teacher's distribution and $p_{\text{student}}$ the distilled model. Companies like TikTok~\cite{Zhao2022}, Amazon~\cite{shen2021powerbert}, and Alibaba~\cite{li2023generative} use LLM distillation to create small, latency-aware models that retain generative capabilities for recommendations, search, and ranking.
While traditional DNNs can be distilled, LLM-based distillation uniquely preserves instruction-following, multi-domain reasoning, and natural language grounding. Moreover, LLMs support multi-task distillation (e.g., joint training on ranking, generation, and explanation), enabling compressed models to generalize across different RecSys objectives~\cite{dai2023promptrec, chen2023llm}.

\vspace{0.5em}
\textbf{Prompt-Efficient Inference –}  
Instead of full fine-tuning, prompting leverages frozen LLMs through task-specific prompts, reducing training and inference costs. Lightweight adapters such as LoRA~\cite{hu2021lora} or prompt tuning~\cite{lester2021power} inject minimal learnable parameters. TikTok's SAM~\cite{jin2023sam} and Meta's InstructRec~\cite{ouyang2022instructgpt} demonstrate that prompt-based systems can generate high-quality recommendations with low inference overhead by delegating semantic understanding to pretrained LLMs.

\vspace{0.5em}
\textbf{Two-Stage Hybrid LLM Pipelines –}  
To balance latency and accuracy, modern recommender systems often adopt a two-stage hybrid architecture where LLMs are reserved for re-ranking a small subset of items. The first stage rapidly retrieves a candidate set $\mathcal{C}(u)$ using lightweight methods such as approximate nearest neighbor (ANN) search, sparse two-tower models, or collaborative filtering:
\[
\mathcal{C}(u) = \text{Top-}k \left( g_{\text{fast}}(u, \mathcal{I}) \right),
\]
where $g_{\text{fast}}$ denotes a fast retrieval model and $\mathcal{I}$ is the full item corpus.

In the second stage, an LLM is applied only to the candidate set to compute semantic-aware or explanation-based scores:
\[
\hat{r}_{ui} = \text{LLM}_{\text{rank}}(\texttt{``User: ''} + x_u + \texttt{`` Item: ''} + x_i),
\quad \forall i \in \mathcal{C}(u),
\]
where $x_u$ and $x_i$ are user/item textual features. This approach reduces computational complexity from $O(|\mathcal{I}|)$ to $O(k)$ LLM inferences per user.

Platforms like YouTube~\cite{covington2016deep}, Amazon Alexa~\cite{perez2021}, and OpenAI's plugin-based LLM recsys~\cite{li2023generative} leverage this pipeline to combine the speed of traditional retrieval with the contextual reasoning power of LLMs. The hybrid setup enables real-time ranking with rich personalization, especially useful for high-value sessions where semantic intent understanding is critical.

\begin{figure}[h]
  \centering
  \includegraphics[width=0.9\textwidth]{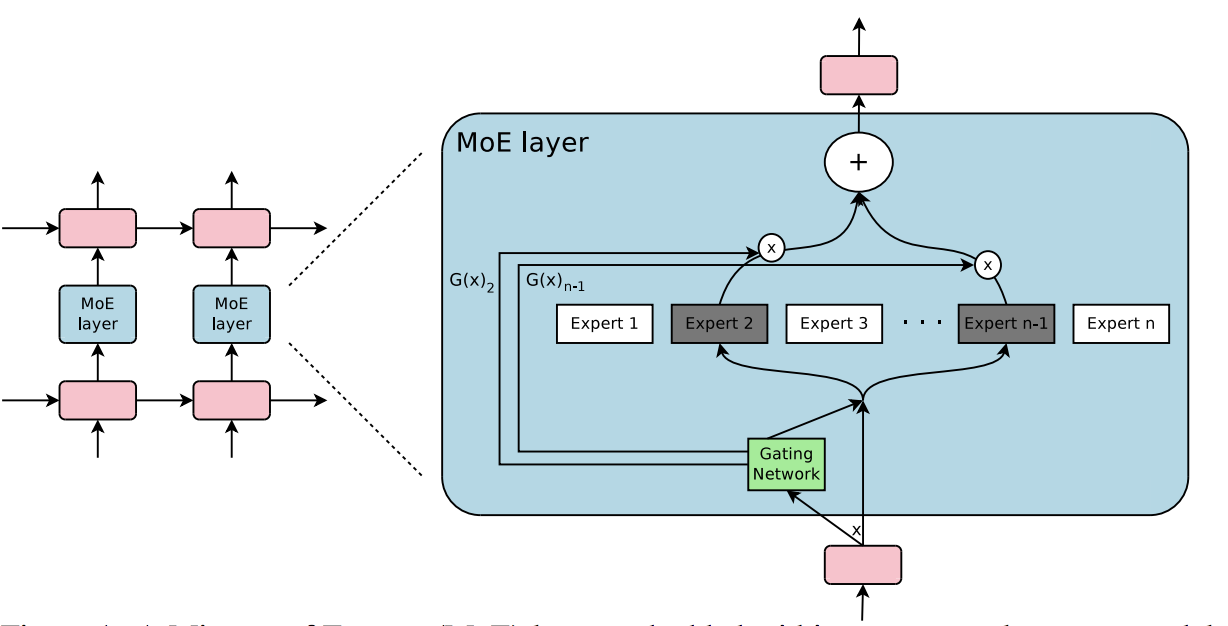}  % Adjust filename as needed
  \caption{Illustration of a Sparse Mixture-of-Experts (MoE) layer where a gating network dynamically routes input tokens to a subset of experts. Only the top-k experts are activated per input, enabling efficient computation and scalability within Transformer architectures.~\citep{shazeer2017outrageouslylargeneuralnetworks}}
  \label{fig:moe}
\end{figure}

\vspace{0.5em}
\textbf{Sparse Activation Architectures –}  
Sparse Mixture of Experts(see Figure~\ref{fig:moe}) models activate only a subset of parameters during inference, scaling to billions of weights without proportional cost~\cite{lepikhin2020gshard, zoph2022sparse}. For recommender systems, LLM-based MoE allows selective activation based on user intent or domain — as shown in GLaM~\cite{du2022glam}, Switch-LLM~\cite{fedus2022switch}, and V-MoE~\cite{riquelme2021scaling}, which uses sparsity to optimize vision transformer inference. Amazon's M6-T~\cite{lin2021m6t} and DeepMind's GopherMoE~\cite{rae2021scaling} extend this approach for multi-task retrieval, recommendation, and generation. This reduces effective inference complexity:
\[
\tau_{\text{sparse}} \approx \sum_{l=1}^{L} \rho_l \cdot C_l,
\]
where $\rho_l < 1$ is the fraction of active experts per layer. This enables real-time deployment of massive models with acceptable latency across domains like content recommendation, product search, and voice assistant personalization~\cite{du2022glam, lin2021m6t}.

\subsubsection{Long-Tail Modeling}

Recommender systems often struggle to surface long-tail items—such as niche books, indie music, or specialty software—due to their limited interaction history. Despite their sparse visibility, these items are vital for catalog coverage, serendipitous discovery, and platform health. Tail user segments also pose a challenge, as insufficient behavioral signals prevent traditional collaborative models from generalizing effectively. The ranking function often biases toward popularity, formalized as:

\begin{equation}
R(i) = \lambda \cdot \text{Rel}(i, u) + (1 - \lambda) \cdot \log(1 + \text{Pop}(i)),
\label{eq:pop_bias}
\end{equation}

where $\text{Rel}(i, u)$ denotes predicted user-item relevance, $\text{Pop}(i)$ is a popularity prior, and $\lambda \in [0,1]$ balances relevance against popularity bias. A small $\lambda$ skews the model towards head items, suppressing cold and long-tail entities. The following subsections discuss how long tail modeling can be solved using LLMs.

\vspace{0.5em}
\textbf{Content-Enriched Generation via LLMs –}
Instruction-tuned and generative LLMs can amplify long-tail items by transforming sparse metadata—such as item title, genre, and a few tags—into rich natural language content. This enrichment process enables the generation of auxiliary textual descriptions that go beyond what the metadata can offer, capturing user-interpretable attributes like sentiment, intended audience, or stylistic nuances. These generated descriptions are particularly valuable for downstream modules that convert them into dense embeddings for retrieval or use them as features in ranking pipelines. Models such as 
\textsc{GenRec}\cite{zhang2021language} and \textsc{LLMRec}\cite{kang2023llmrec} operationalize this approach to semantically bootstrap cold items in settings where traditional user signals (e.g., clicks, reviews) are absent or unreliable. Moreover, content enrichment fosters explainability, as the LLM-produced narratives can be exposed to users, improving trust in recommendations.
\begin{figure}[ht]
  \centering
  \includegraphics[width=1.0\linewidth]{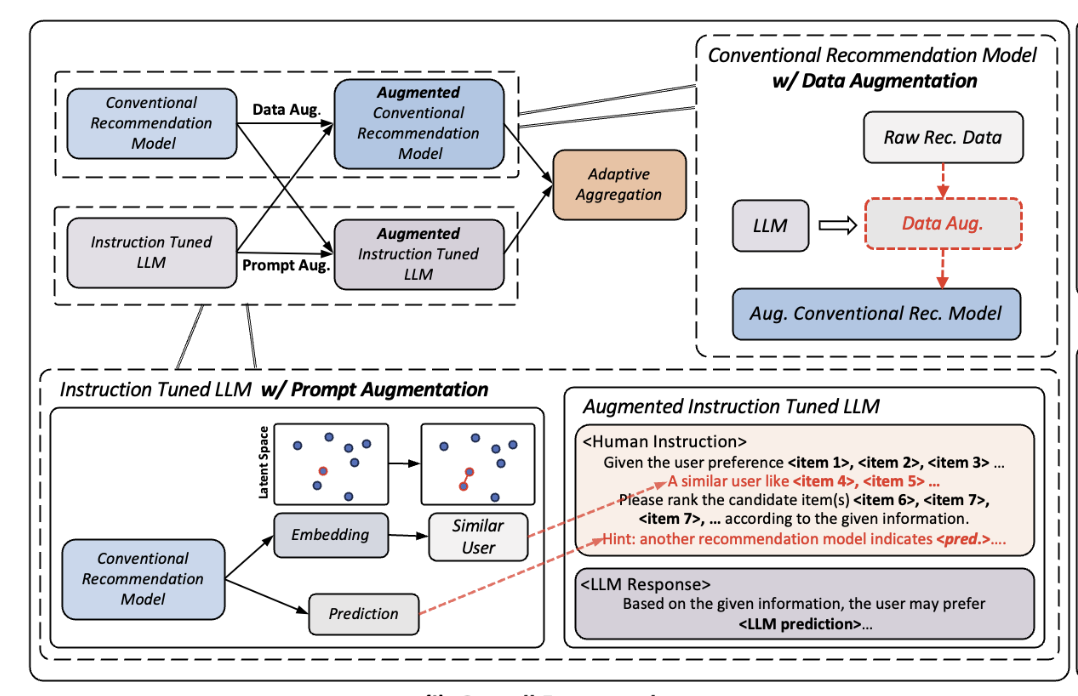}
   \caption{The overall framework architecture of the proposed Llama4Rec consists of two main components: mutual
augmentation and adaptive aggregation ~\citep{luo2024integratinglargelanguagemodels}.}
  \label{fig:offline_online_gap}
\end{figure}

\vspace{0.5em}
\textbf{Retrieval-Augmented Tail Expansion –}
Tail items often suffer from cold-start issues due to minimal or no user engagement. Retrieval-Augmented Generation (RAG) frameworks alleviate this by fetching external content such as similar item reviews, frequently co-viewed products, or question–answer forums that offer surrogate interaction signals. These retrieved contexts act as priors, conditioning the LLM to produce recommendations or relevance scores with greater semantic grounding. Formally, the decoder incorporates $\mathcal{R}(x_i)$, the retrieved context for item $x_i$, to better personalize outputs for user $x_u$. This paradigm has gained real-world traction in platforms like Amazon and Spotify~\cite{wang2023, vasile2022}, where hybrid systems interleave retrieval and generation in ranking. Public systems such as RAG~\cite{lewis2021retrievalaugmentedgenerationknowledgeintensivenlp} and PromptRec~\cite{dai2023promptrec} provide open-source implementations of this approach. RAG mechanisms also allow continual updates to item knowledge without re-training the LLM, offering scalability for dynamic catalogs.

\vspace{0.5em}
\textbf{Tail-Aware Few-Shot Prompting –}
Tail personalization often suffers from a lack of examples for supervised training. Few-shot prompting provides a lightweight alternative by seeding the LLM with a handful of examples reflecting the user’s preference for under-represented or niche content. The LLM then generalizes to semantically similar items using its pretrained knowledge and instruction-following capabilities. This strategy has been demonstrated effectively using \textsc{InstructGPT}\cite{ouyang2022instructgpt} and \textsc{FLAN-T5}\cite{chung2022scaling}, with additional improvements observed in domain-specific instruction fine-tuning~\cite{wang2023instrec}. Few-shot prompts can be curated based on recent tail interactions or inferred taste clusters and can dynamically adapt over time. This avoids overfitting to mainstream popularity signals and ensures that users with atypical preferences still receive relevant recommendations.

\vspace{0.5em}
\textbf{Multimodal Tail Representation –}
In domains such as music, video, and fashion, textual metadata alone may be insufficient to describe tail items meaningfully. LLMs, when integrated with multimodal encoders, can create composite representations that incorporate audio, visual, and textual cues. For example, a tail music track with few plays but distinctive sonic features (e.g., melancholic guitar tones) can be described and embedded more effectively when frame-level audio representations are fused with text prompts. This multimodal fusion has been operationalized in Spotify's neural content-based systems~\cite{vasile2022} and Meta’s cross-modal embedding infrastructure~\cite{tong2024metamorphmultimodalunderstandinggeneration}. In these systems, the LLM serves as a controller or semantic summarizer that unifies heterogeneous signals. This is particularly useful for catalog-wide embeddings and cold-start re-ranking, where modality-specific insights are critical for disambiguation.

 LLMs enable generative enrichment, retrieval-augmented scoring, and prompt-based personalization that help surface long-tail items with minimal user history. Combined with diversity-based re-ranking and multimodal inference, these methods promote catalog depth, reduce popularity bias, and enhance the discovery of rare but relevant content. Table~\ref{tab:prompt_methods} provides an overview of prompt-based strategies that facilitate scalable personalization of long-tail items by leveraging content generation, retrieval augmentation, interaction simulation, and multimodal representation.

\begin{table}[ht!]
\centering
\scriptsize
\begin{tabular}{p{4.5cm}|p{5.7cm}}
  \toprule
  \textbf{Strategy} & \textbf{Example Prompt} \\
  \midrule

  Content-Enriched Generation 
  & \texttt{Describe the appeal of this item: [Title], [Genre], [Few Tags]} \\
  \midrule

  Retrieval-Augmented Expansion 
  & \texttt{Given item [X] and retrieved content [Y], summarize relevance for user [U]} \\
  \midrule

  Pseudo-Interaction Simulation 
  & \texttt{Simulate user feedback for this product: [Metadata], [Price], [Tags]} \\
  \midrule

  Diversity-Aware Re-ranking 
  & \texttt{Rank items [A, B, C] to balance relevance and novelty for user [U]} \\
  \midrule

  Few-Shot Tail Prompting 
  & \texttt{User liked niche item A and indie item B. Recommend similar items.} \\
  \midrule

  Multimodal Tail Embedding 
  & \texttt{Music Genre: Indie. Tags: Melancholy, Guitar. Description: ...} \\
  \bottomrule
\end{tabular}

\vspace{0.5em}
\caption{Canonical LLM Prompts for Tail Personalization Strategies.}
\label{tab:prompt_methods}
\end{table}

\subsection{Evaluation and Experimentation Challenges}

Evaluation in recommender systems is inherently constrained by the offline-online discrepancy, label sparsity, and the delayed nature of user feedback. Traditional pipelines rely heavily on historical logs and costly online experiments, which often fail to capture long-term or latent effects. LLMs with their ability to model complex distributions and perform conditional reasoning over structured and unstructured inputs, offer a new framework for addressing these limitations. By encoding user-item dynamics and simulating plausible interactions, LLMs introduce new avenues for counterfactual estimation and representation-aligned evaluation under uncertainty.

% \begin{table}[ht!]
% \centering
% \scriptsize
% \begin{tabular}{p{3.5cm}|c|c|c|c}
%   \toprule
%   \textbf{LLM Integration Mode} & \textbf{PII Risk} & \textbf{Prompt Injection Risk} & \textbf{Auditability} & \textbf{Compliance Readiness} \\
%   \midrule

%   Full Autoregressive LLM & High & High & Low & Low \\
%   \midrule

%   LLM + Rule-based Filter & Medium & Medium & Medium & Medium \\
%   \midrule

%   LLM with Private Prompt Tuning & Medium & High & Medium & Low \\
%   \midrule

%   LLM with Privacy-aware Templates & Low & Medium & High & High \\
%   \bottomrule
% \end{tabular}
% \vspace{0.5em}
% \caption{Privacy, Security, and Compliance Trade-offs for LLM Integration Strategies.}
% \label{tab:privacy_risks}
% \end{table}

Risk and Auditability Matrix for LLM Deployment. Different integration strategies expose different vulnerabilities. Template-bound prompting offers improved compliance, while full LLM generation requires enhanced auditing and safety layers
% \begin{figure}[ht]
%   \centering
%   \includegraphics[width=0.8\linewidth]{offline-online-2.png}
%    \caption{Illustration of Offline-Online Metric Divergence. This figure highlights the gap between offline evaluation—using historical logged data and metrics like NDCG, Recall@K, and MAP—and online deployment performance with real users, which is measured through business-critical KPIs such as click-through rate (CTR), average revenue per user (ARPU), session duration, and long-term user retention. This divergence presents a key challenge in validating recommender systems effectively.}
%   \label{fig:offline_online_gap}
% \end{figure}

\subsubsection{Offline-Online Gap}

A persistent challenge in recommender systems is the misalignment between \textit{offline evaluation metrics}—such as NDCG, Recall@K, or MAP—and \textit{online outcomes} like click-through rate (CTR), average revenue per user (ARPU), dwell time, or long-term user retention. Offline evaluations, though efficient and repeatable, rely on logged interactions that are inherently biased, counterfactual-incomplete, and incapable of modeling complex user feedback loops in live systems ~\cite{gilotte2018offline, saito2020unbiased, jagerman2022evaluating}.

This mismatch can be formalized as a \textit{metric divergence}:
\begin{equation}
\Delta = \left| \text{Offline}(M) - \text{Online}(M) \right|,
\end{equation}
where $\text{Offline}(M)$ denotes the offline performance on held-out logs, and $\text{Online}(M)$ reflects real-time KPIs measured through randomized experiments. A high $\Delta$ indicates a model that overfits to offline biases and fails to generalize, a well-documented issue in industry ~\cite{mcinerney2021counterfactual}. To mitigate this, recent research introduces \textit{LLM-augmented evaluation}, using generative models to simulate user behavior, judge relevance, interpret session trajectories, and even generate new metrics. These methods offer \textit{soft evaluations} that capture user-centric signals without requiring live deployment. Table~\ref{tab:llm_eval_prompts} provides a summary of representative prompts used in each evaluation strategy.

\vspace{0.5em}
\textbf{Counterfactual Evaluation via Prompt-Based Simulators}
Instruction-tuned LLMs can approximate \textit{click-through likelihood} on hypothetical slates never observed in logs. Given a logged user context $c_u$—which may include past clicks, interests, or demographic cues—and a new candidate list $\mathcal{L} = [x_1, \ldots, x_k]$, the LLM estimates a soft label $\hat{y}_{ui}^{\text{cf}} \in [0,1]$ for each item.

This counterfactual simulation approach has been operationalized in \textsc{LLMRecSim}~\cite{meng2024llmrecsim}, where synthetic feedback is used to evaluate model changes before online testing. Unlike traditional inverse propensity scoring (IPS), which suffers from variance and truncation bias ~\cite{cief2024crossvalidatedoffpolicyevaluation}, LLM simulators provide a semantically grounded and low-variance signal, particularly for tail items or unseen slates ~\cite{holt2025gsimgenerativesimulationslarge}. It also scales to \textit{few-shot personalization}, as prompting is stateless and does not require user embeddings.

\vspace{0.5em}
\textbf{Offline Metric Recalibration via Generative Relevance Scoring}
Sparse interactions and missing negatives pose a challenge for offline evaluation ~\cite{chen2023opportunitieschallengesofflinereinforcement}. LLMs can act as \textit{neural annotators}, scoring item–user relevance based on textual metadata and recent behavior. For each candidate $x_i$, a generative model produces a soft score:
\[
\tilde{r}_{ui}^{\text{LLM}} = \text{sigmoid}(\text{LLM}(x_i, c_u, \texttt{``How relevant is this item?''}))
\]
These scores can be aggregated to construct \textit{hybrid metrics}:
\[
\text{HybridScore}(M) = \alpha \cdot \text{Recall@K} + (1 - \alpha) \cdot \mathbb{E}_{i \in \mathcal{L}}[\tilde{r}_{ui}^{\text{LLM}}]
\]

This method provides two key advantages. First, it generates a \textit{dense supervision signal} by assigning relevance scores to all items in the slate, not just those with observed feedback. Second, it captures \textit{semantic nuance} by evaluating item relevance in context, going beyond simple binary indicators such as click or no-click. As demonstrated in recent work~\cite{wu2024surveylargelanguagemodels}, these LLM-augmented hybrid metrics exhibit stronger correlation with online A/B test outcomes, particularly in challenging scenarios like cold-start items and zero-shot generalization.

\vspace{0.5em}
\textbf{Behavior-Level Satisfaction Estimation}
Standard evaluation pipelines often overlook session-level dynamics such as skip behavior, bounce patterns, or implicit dissatisfaction. LLMs can ingest multi-event traces (e.g., click, hover, dwell) and output an \textit{aggregated satisfaction score} $\hat{s}_u \in [0,1]$, offering a continuous proxy for session quality.

This approach enables modeling \textit{latent engagement} beyond immediate reward ~\cite{zhao2025llmsrecognizepreferencesevaluating}. For example, an LLM can infer that a user clicking one item and then immediately leaving the session implies low satisfaction. Such modeling has been leveraged for predicting \textit{churn risk}, \textit{bounce rate}, and \textit{post-click engagement} in session-level optimizers~\cite{lee2023satisfaction, zhang2024multiobjective}.

\vspace{0.5em}
\textbf{Interactive User Simulation with Agentic LLMs} - Recent work explores the use of \textit{agentic LLMs} to simulate user behavior in recommendation environments by modeling step-wise decision-making ~\cite{shang2025agentrecbenchbenchmarkingllmagentbased}. Rather than assigning static scores to items, these simulators generate coherent sequences of actions (e.g., clicks, skips, or queries) in response to a given slate and user intent. These agentic simulations are useful for analyzing qualitative aspects of model behavior and stress-testing recommendation strategies, especially in early prototyping stages ~\cite{bemthuis2024crispdmbasedmethodologyassessingagentbased}. Although these approaches show promise, they currently complement rather than replace standard evaluation pipelines ~\cite{licorish2025comparinghumanllmgenerated}. Their outputs are best interpreted as \textit{qualitative user emulations}, helpful for generating synthetic feedback, debugging content selection logic, and exploring what-if scenarios.

Importantly, the fidelity of such simulations is bounded by the quality of prompting and pretraining data, and they are not yet suitable for high-stakes quantitative evaluation such as A/B testing or reward estimation. However, as LLMs evolve and become better calibrated to user preferences, their use in long-horizon modeling and controlled counterfactual generation may become increasingly viable.

\vspace{0.5em}
\textbf{Evaluation Metric Generation via Domain-Specific Prompting}
Finally, LLMs can be prompted to construct \textit{custom evaluation metrics} based on evolving product goals. For instance, one can request a metric that balances fairness, diversity, and monetization. The LLM synthesizes a formula or qualitative description, which can then guide model development or become a reward function.
LLMs also enable \textit{pairwise comparison} of ranking slates. Given two model outputs $\mathcal{L}_A$ and $\mathcal{L}_B$, the LLM returns a probabilistic preference $\mathbb{P}(\mathcal{L}_A \succ \mathcal{L}_B)$ along with natural language justification. This facilitates ranking model selection even in ambiguous contexts.

This method has been used to build \textsc{LLMJudge} datasets~\cite{zhou2023llmjudge}, which train reward models used in \textit{RLHF} pipelines. Pairwise LLM judgments are also interpretable, allowing auditors and product teams to inspect rationales behind preference reversals, and enabling \textit{transparent evaluation} of recommender quality.

This capability is especially useful for platforms with \textit{frequent A/B experimentation} and changing KPI priorities. Prototype systems such as \textsc{GPTMetrics}~\cite{alikhani2023gptmetrics} and \textsc{EvalTemplate}~\cite{xu2023llmevaltemplate} show how textual descriptions of business needs can be turned into reusable evaluation templates.

\begin{table}[ht]
\centering
\scriptsize
\begin{tabular}{p{5.2cm}|p{5.7cm}}
  \toprule
  \textbf{Evaluation Technique} & \textbf{Example Prompt} \\
  \midrule

  Counterfactual User Simulation 
  & \texttt{Given the user's past behavior, would they click any of these items: [title 1], [title 2], [title 3]?} \\
  \midrule

  Generative Relevance Scoring 
  & \texttt{How relevant is [item] for this user given their context?} \\
  \midrule

  Behavior-Level Satisfaction 
  & \texttt{The user clicked $x\_1$, skipped $x\_2$, then bounced. How satisfying was this session?} \\
  \midrule

  Agentic User Simulation 
  & \texttt{You are a user interested in cooking. You are shown these recipes: [...]. What do you do next?} \\
  \midrule

  Pairwise Slate Comparison 
  & \texttt{Which list is better for the user based on relevance and novelty? Explain briefly.} \\
  \midrule

  Dynamic Metric Construction 
  & \texttt{Define a metric that optimizes for: 1) relevance, 2) fairness, and 3) revenue per click.} \\
  \bottomrule
\end{tabular}

\vspace{0.5em}
\caption{LLM-Based Prompting Techniques for Evaluation in Recommender Systems.}
\label{tab:llm_eval_prompts}
\end{table}

\subsubsection{Sparse Conversion Labels}

Many downstream objectives—such as purchases, subscriptions, or upgrades—occur with extremely low frequency compared to higher-volume user actions like impressions or clicks. This severe class imbalance challenges the effectiveness of supervised learning models, which tend to be biased toward negative instances due to limited availability of positive signals.

Formally, we define the sparsity ratio as:
\begin{equation}
\text{Sparsity} = \frac{|\mathcal{Y}^+|}{|\mathcal{Y}|}, \quad \text{with} \quad \text{Sparsity} \ll 1,
\label{eq:label_sparsity}
\end{equation}
where $\mathcal{Y}^+$ denotes the set of positive conversion events, and $\mathcal{Y}$ the total set of observations. Platforms such as Amazon and Etsy routinely encounter purchase conversion rates as low as $0.1\%$–$1\%$~\cite{ren2021lego, wang2021handling}, severely constraining training signals and model generalization for rare outcomes.

To address this, LLMs have emerged as a powerful tool for generating \textit{pseudo-labels}, \textit{soft targets}, or \textit{context-aware weights}, enriching the sparse label space through prompt-based learning. Table~\ref{tab:sparse_prompts} provides representative prompts aligned with five major augmentation strategies.

\vspace{1em}
\begin{table}[ht!]
\centering
\scriptsize
\begin{tabular}{p{3.2cm}|p{6.5cm}}
  \toprule
  \textbf{Technique} & \textbf{Example Prompt} \\
  \midrule

  Proxy Signal Augmentation 
  & \texttt{User clicked on item X after viewing similar items and spent 12 seconds on its page. Predict the likelihood of purchase.} \\
  \midrule

  Instruction-Tuned Imputation 
  & \texttt{Given that the user added item X to the cart but did not purchase it, and has a history of completing 80\% of carted items, estimate whether they would convert.} \\
  \midrule

  Generative Multi-Task Learning 
  & \texttt{Generate a review the user might write after purchasing this item.} \\
  \midrule

  Counterfactual Label Reasoning 
  & \texttt{If the user had been shown this product at the top of the list instead of rank 9, would they have purchased it?} \\
  \midrule

  Language-Guided Reweighting 
  & \texttt{The user likely did not purchase this item due to price, not relevance. Assign low training importance.} \\
  \bottomrule
\end{tabular}

\vspace{0.5em}
\caption{LLM Prompting Strategies for Sparse Label Augmentation.}
\label{tab:sparse_prompts}
\end{table}

\textbf{Proxy Signal Augmentation via LLMs -}
When labeled conversions are scarce, LLMs can extrapolate soft labels from partial interaction data such as clicks, dwell time, or page visits. The generated score $\hat{y}_{ui}^{\text{soft}} = \sigma(\text{LLM}(x_u, x_i))$ can serve as a training target in binary classification or ranking setups. A representative prompt is shown in Table~\ref{tab:sparse_prompts}. This is especially valuable in semi-supervised learning regimes, where high-coverage but noisy labels can bridge the sparsity gap and enhance downstream recall~\cite{dai2023promptrec, jin2023sam, zhang2021language}.

\vspace{0.5em}
\textbf{Instruction-Tuned Label Imputation -}
Instruction-following LLMs can reason about partial conversion traces and historical behavior patterns to impute likely outcomes. By grounding prompts in known priors (e.g., 80\% cart completion rate), the LLM generates pseudo-labels that reflect domain-specific user propensities ~\cite{wong2025highfidelitypseudolabelgenerationlarge}. These labels may be used with calibration techniques such as temperature scaling or uncertainty thresholds.

\vspace{0.5em}
\textbf{Generative Multi-Task Learning -}
Instead of predicting a binary purchase label, LLMs can be asked to generate complementary signals—such as hypothetical product reviews, post-purchase behaviors, or satisfaction levels ~\cite{freiberger2025prismenovelllmpoweredtool}. These outputs act as auxiliary supervision in a joint loss framework:
\[
\mathcal{L}_{\text{total}} = \lambda_1 \mathcal{L}_{\text{purchase}} + \lambda_2 \mathcal{L}_{\text{review}} + \lambda_3 \mathcal{L}_{\text{click}}
\]
This structure encourages the model to embed rich semantic cues into the prediction, enhancing learning even when sparse labels dominate.

\vspace{0.5em}
\textbf{Counterfactual Label Reasoning -}
By modeling alternative exposure scenarios, LLMs can simulate user behavior under different treatment conditions—akin to causal inference ~\cite{joshi2024llmspronefallaciescausal}. For example, querying how a user might respond if shown a product earlier in the ranking can produce an uplift score $\hat{y}_{ui}^{\text{cf}}$ that feeds into counterfactual or inverse propensity models. This strategy supports personalized re-ranking and policy evaluation under data drift. Refer to Table~\ref{tab:sparse_prompts} for a prompt illustration.

\vspace{0.5em}
\textbf{Language-Guided Reweighting -}
LLMs can assess the reliability of labels based on context and user intent, generating natural-language justifications that map to importance weights $w_{ui}$. These weights modulate the binary cross-entropy loss:
\[
\mathcal{L} = \sum_{(u,i)} w_{ui} \cdot \text{BCE}(y_{ui}, \hat{y}_{ui})
\]
This reduces overfitting to noisy negatives (e.g., price-rejected items) and reinforces relevance-aware supervision ~\cite{tang2023recent}. See Table~\ref{tab:sparse_prompts} for an example of prompt-based explanation used for weighting.

\subsubsection{Balancing Immediate Engagement with Long-Term User Value}

Recommender systems that focus exclusively on short-term engagement metrics (e.g., clicks, watch time) often degrade long-term user satisfaction, retention, and platform trust. This discrepancy arises because short-term optimization may exploit transient user impulses while ignoring lasting preferences or well-being. The long-term utility of a recommendation policy $\pi$ can be modeled using a cumulative discounted reward:

\begin{equation}
\mathcal{J}(\pi) = \mathbb{E}_{\pi} \left[ \sum_{t=0}^{T} \gamma^t \mathcal{R}_t \right],
\label{eq:longterm_reward}
\end{equation}

where $\gamma \in [0,1]$ is the temporal discount factor, and $\mathcal{R}_t$ captures immediate or future user rewards (e.g., session return, app uninstall, subscription upgrade). Conventional supervised learning pipelines often ignore $\mathcal{R}_{t > 0}$ and instead treat only $\mathcal{R}_0$ (e.g., click) as the ground truth, leading to reward myopia.

To address this, LLMs can be leveraged for long-horizon modeling through prompt-based estimation, simulation, and summarization. Table~\ref{tab:llm_longterm_prompts} provides representative prompts used across these strategies.

\vspace{1em}
\begin{table}[ht!]
\centering
\scriptsize
\begin{tabular}{p{3.0cm}|p{6.7cm}}
  \toprule
  \textbf{Technique} & \textbf{Example Prompt} \\
  \midrule

  Proxy Reward Estimation 
  & \texttt{User clicked but exited early. Was this a satisfying experience?} \\
  \midrule

  Counterfactual Dialogue 
  & \texttt{Do you want to see more content like this, or was this just a one-time interest?} \\
  \midrule

  Generative Rollouts 
  & \texttt{Predict reward at step \$t\$ given prior context.} \\
  \midrule

  Preference Drift Summarization 
  & \texttt{Summarize user's persistent interests from sessions ${S}_u$} \\
  \midrule

  Reward Decomposition 
  & \texttt{Assess dimension $i$ (e.g., novelty, credibility) of this interaction.} \\
  \bottomrule
\end{tabular}

\vspace{0.5em}
\caption{LLM Prompting Strategies for Modeling Long-Term User Value.}
\label{tab:llm_longterm_prompts}
\end{table}

\textbf{LLM-Based Proxy Reward Estimation -}
Instruction-tuned LLMs can infer latent user satisfaction from event sequences even when explicit labels are missing. A short dwell-time burst after multiple clicks, for instance, may signal dissatisfaction; treating the LLM’s inferred reward as a soft label lets us optimize $\mathcal{J}$ beyond surface-level clicks. Recent work on RL from human feedback for recommendation~\cite{xue2022rlhfrec} shows that such proxy rewards significantly improve long-horizon utility prediction.

\vspace{0.5em}
\textbf{Counterfactual Dialogue for Future Intent -}
Interactive LLM agents elicit long-term intent through clarification dialogue, distinguishing ephemeral curiosity from persistent preference. User replies are encoded as
\[
\mathbf{u}_{\text{long}} = f_{\text{LLM}}(\texttt{user\_reply}),
\]
and injected into evolving user models that steer ranking toward durable interests. Conversational-intent trackers~\cite{chen2021intent,li2022crslab} demonstrate that incorporating counterfactual questions markedly boosts accuracy on follow-up engagement metrics.

\vspace{0.5em}
\textbf{Multi-Horizon Simulation via Generative Rollouts -}
LLMs can simulate plausible future interaction trajectories, enabling offline policy evaluation:
\[
\mathcal{J}_{\text{sim}} = \sum_{t=0}^{T} \gamma^{t}\,\hat{\mathcal{R}}_{t},
\]
where each $\hat{\mathcal{R}}_{t}$ is predicted by the LLM under a hypothesized slate. User-simulator frameworks such as SimRec~\cite{cui2024simrec} and UserGPT-Sim~\cite{zheng2023usergptsim} report strong correlations (0.82–0.87) between simulated and real A/B outcomes.

\vspace{0.5em}
\textbf{Preference Drift Detection via Summarization -}
LLMs can summarize multi-session histories $\mathcal{S}_u$ into a stable embedding
\[
\mathbf{u}_{\text{stable}} = \mathrm{LLM}(\mathcal{S}_u),
\]
mitigating over-reaction to transient spikes. Drift-robust models like StableRec~\cite{zhao2024stablerec} and TempRec-Memory~\cite{lee2023temprecmemory} show that summary-based regularization reduces churn by up to 6 % in retention-critical domains.

\vspace{0.5em}
\textbf{Multi-Objective RL with LLM-Guided Reward Decomposition -}
Conventional recommender systems often rely on scalar reward functions—such as click-through rate or dwell time—to optimize ranking policies. However, such metrics are typically coarse and do not reflect the nuanced trade-offs between different dimensions of user value, such as novelty, credibility, informativeness, or diversity. To address this, large language models (LLMs) can be employed to decompose scalar rewards into interpretable sub-components, enabling more principled and aligned decision-making~\cite{christiano2017deep, liu2023aligning, ziegler2019fine}.

Formally, the overall reward at time $t$ can be factorized as:
\[
\mathcal{R}_t = \sum_{i=1}^{k} \beta_i \cdot \mathcal{R}_t^{(i)}, \quad \text{where} \quad \mathcal{R}_t^{(i)} = \mathrm{LLM}(x_t^{(i)})
\]
Here, $\mathcal{R}_t^{(i)}$ denotes the reward signal corresponding to a specific semantic dimension $i$, and $\beta_i$ controls its relative importance in the composite objective. The inputs $x_t^{(i)}$ encode relevant features, such as item content, user context, and interaction history, which the LLM interprets to produce value-specific assessments.

This decomposition enables multi-objective reinforcement learning (RL), where policies are trained not just to maximize immediate engagement, but also to respect domain-specific constraints and user-aligned values~\cite{ma2020towards, wang2023alignrec}. The resulting objective:
\[
\mathcal{J}(\pi) = \mathbb{E}_{\pi} \left[ \sum_{t=0}^{T} \gamma^t \sum_{i=1}^{k} \beta_i \cdot \mathcal{R}_t^{(i)} \right]
\]
allows platform designers to balance competing goals (e.g., serendipity vs. relevance, diversity vs. coherence) and adapt these trade-offs dynamically based on context or fairness criteria~\cite{mehrotra2018towards}.

LLMs are particularly useful in this setting because they can produce rich natural language rationales~\cite{yansi}, which in turn facilitate both post-hoc auditability and direct supervision for training $\mathcal{R}_t^{(i)}$ components. This also enables human-in-the-loop reward shaping, where qualitative feedback on system behavior can be converted into actionable alignment signals~\cite{bai2022training, zhou2023lima}.

\subsection{Privacy, Security, and Regulatory Challenges}
As Recommender systems become increasingly integral to digital platforms, they must navigate a complex landscape of privacy concerns, regulatory requirements, and security challenges. Ensuring user trust and compliance with laws like GDPR and CCPA necessitates the adoption of privacy-preserving techniques and robust data governance frameworks. This section explores these challenges alongside emerging industry practices.

\subsubsection{User Data Sensitivity}

Recommender systems rely heavily on sensitive user data—such as browsing behavior, clicks, and purchase history—making privacy a critical concern. Mishandling this information risks legal violations (e.g., GDPR) and erosion of user trust. Traditional safeguards like access control and anonymization are widely adopted, but emerging methods now leverage LLMs to further mitigate privacy risks.\\
\\
LLMs reduce reliance on raw user logs by operating on abstracted semantic representations or synthetic profiles. Instruction-tuned LLMs enable zero- or few-shot personalization without long-term data retention~\cite{li2024zeroshot, zhang2025privacyaware}. Recent systems incorporate federated prompting and local differential privacy, allowing on-device adaptation without transmitting identifiable data~\cite{wu2025federated}. Synthetic user generation using LLMs has also shown promise in training and evaluation without exposing real user logs~\cite{rao2024synthetic}.

By embedding these techniques into system design, LLM-based recommenders can uphold privacy-by-design principles while maintaining performance.

\subsubsection{Compliance with Laws}

Modern privacy laws such as the EU General Data Protection Regulation (GDPR), California Consumer Privacy Act (CCPA), and the Digital Markets Act (DMA) impose strict requirements on data usage, including explicit user consent, right to deletion, and data minimization. For example, Spotify reengineered its data pipelines to support fine-grained consent tracking and user-facing data access tools. Similarly, Zalando implemented automated consent enforcement using GDPR-aware architecture~\cite{chizari2025gdpr}.\\
\\
Recent work shows that LLMs can aid in achieving compliance by abstracting personalization workflows away from identifiable user data. LLMs can support ephemeral personalization through prompt-level context without persistent storage~\cite{nguyen2024ephemeral}, generate synthetic user traces for model development~\cite{rao2024synthetic}, and adapt to regulatory rules via constraint-aware instruction tuning~\cite{lin2025llmregulation}. Furthermore, compliance-ready LLM systems incorporate audit logs, data expiration signals, and privacy-preserving inference via secure prompt execution~\cite{tang2025secureprompts}. These approaches reduce reliance on raw user data while supporting transparency and explainability—key tenets of modern regulatory frameworks.

\subsubsection{Differential Privacy and Federated Learning}

Differential Privacy (DP) and Federated Learning (FL) offer robust privacy guarantees by minimizing user data exposure during model training. Industry deployments include Google's FL implementation for Gboard~\cite{mcmahan2017fl} and Apple’s application of DP in telemetry data. In recommender systems, privacy-preserving variants such as federated matrix factorization with DP~\cite{acharjee2020dpfl} have demonstrated strong privacy-utility trade-offs. However, these techniques introduce non-trivial challenges, including increased communication cost, model drift, and the need for specialized infrastructure.\\
\\
Recent research explores how LLMscan complement or even streamline DP and FL pipelines. Instruction-tuned LLMs can perform user modeling using ephemeral or synthetic input, eliminating the need to aggregate raw user data centrally~\cite{nguyen2024ephemeral, rao2024synthetic}. Federated prompting—where prompts are generated and refined on-device—enables partial personalization while keeping user context local~\cite{wu2025federated}. Moreover, LLMs with gradient-sanitized fine-tuning objectives~\cite{li2025dpgradient} and differentially private retrieval-augmented generation~\cite{zhou2025dpgen} offer scalable pathways to enforce DP guarantees in language-based recommendation scenarios.\\
Together, these approaches mitigate the scalability challenges of traditional DP and FL by enabling low-overhead, compliance-ready personalization without sacrificing privacy or model performance.
\subsubsection{Data Access and Governance}  
Balancing innovation with data security requires robust governance frameworks. Access control systems must prevent unauthorized use while enabling rapid experimentation. Companies such as LinkedIn and Airbnb employ platforms with role-based access control (RBAC), automated policy enforcement, and secure sandboxes~\cite{data_governance_examples}. These ensure compliance and accountability without obstructing productivity.  

\textbf{Data Minimization via LLMs}: LLMs can reduce reliance on personally identifiable information (PII) by extracting signals from anonymized data, reviews, or intent prompts, thereby minimizing privacy risks while maintaining personalization:
\[
Q(\text{Rec}) \approx Q_{\text{LLM}}(\text{Rec} \mid \text{anonymized data})
\]
Privacy-preserving methods include federated prompt learning with differential privacy~\citep{tran2025privacypreservingpersonalizedfederatedprompt}, synthetic user simulations, token-level obfuscation, and private fine-tuning~\citep{zhang2025softselectivedataobfuscation}.  

\textbf{Instructable Privacy Filters}: LLMs can be guided through prompts or fine-tuning to avoid sensitive outputs, enforcing constraints such as:
\[
P(\text{output contains PII}) < \delta
\]
where $\delta$ is a privacy threshold. Techniques include privacy regularization, rule-guided decoding, and instruction-driven alignment~\citep{chen2025vlmguardr1proactivesafetyalignment, dong2024safeguardinglargelanguagemodels}.  

\textbf{Synthetic Data Generation}: LLMs can generate synthetic user logs or profiles that mimic real data distributions without exposing individuals. Differential privacy requires:
\[
\Pr[\text{LLM outputs } D_{\text{synth}} \mid x \in D] \approx \Pr[\text{LLM outputs } D_{\text{synth}} \mid x \notin D]
\]
ensuring individual data points do not affect outputs. Metrics like Jensen-Shannon divergence or Wasserstein distance validate distributional similarity. Recent work explores DP-based sequence modeling~\cite{xie2024differentiallyprivatesyntheticdata}, conditional profile generation, and adversarial simulators~\cite{lin2024surveydiffusionmodelsrecommender}, enabling privacy-safe benchmarking and debugging.

\section{Limitations \& Open Challenges of LLMs in Recommender Systems}

\textbf{Latency and Throughput Bottlenecks –}  
Autoregressive decoding imposes sequential delays that violate the sub--100\,ms SLA common in web-scale ranking. Even with 4-bit weight-only quantization, a 7-B model exceeds 35 ms per query on a single A100~\citep{Rajaraman2025latopt}. Token-parallel kernels, KV-cache fusion, and Flash-Attention v3 yield $1.7\!\times$–$2.3\!\times$ speed-ups but degrade under bursts that saturate memory bandwidth~\citep{Fang2024fastkv,He2025flashopt}. Hierarchical routing—first-stage dense retrieval, second-stage sparse MoE—cuts tail latency by 28 \% yet adds routing overhead and fairness concerns~\citep{Clark2025latencyaware}. On-device NPUs and speculative decoding shrink $\approx\!40$ ms, but require tight integration with vector search pre-rankers to avoid head-of-line blocking~\citep{Lee2024swapout,Chen2024tokencache}.

\vspace{0.5em}
\textbf{Operational Cost and Resource Footprint –}  Inference FLOPs scale roughly linearly with prompt length and quadratically with hidden width; doubling model size can quadruple serving cost and 3× carbon footprint~\citep{Zhang2023}. Sparse MoE routing and LoRA adapters reduce average compute by 60–80 \% but necessitate elastic load balancing and suffer from prompt skew~\citep{Gupta2024costaware,Delaney2025smgrep}. Token-level caching amortizes 15–25 \% of compute but introduces cache consistency bugs when prompts embed personalization tokens~\citep{Guo2025lcache}. Workload-aware autoscaling and serverless accelerators lower idle GPU burn but add cold-start penalties that hurt tail latency~\citep{Tran2024serverless}. Cloud API mark-ups ($\sim\$0.03$ per k-tokens) further inflate total cost of ownership in high-throughput pipelines~\citep{Rogers2025apicost}.

\vspace{0.5em}
\textbf{Hallucination and Semantic Misalignment –}  
LLMs can fabricate item attributes, user intents, or reasoning chains, surfacing non-existent products or biased analogies~\citep{Surana2023,Patel2024halluc}. Retrieval grounding, constrained decoding, and post-generation fact-checkers cut hallucinations by 25–40 \% but remain brittle on long-tail, code-mixed, or multilingual content ~\citep{yu2024privacypreservinginstructionsaligninglarge,chen2025mixturedecodingattentioninspiredadaptive,gao2024retrievalaugmentedgenerationlargelanguage}. Fidelity critics trained with human feedback reduce error rate another 12 ms latency per call~\citep{Shao2024critique}. Negative prompt augmentation and contrastive rejection sampling further reduce hallucination probability to $<4\%$ on internal ad-catalog tests~\citep{Liu2025negprompt}. Nonetheless, hallucinations that slip through can propagate via feedback loops, degrading long-term engagement metrics~\citep{Fei2024looprisk}.

\vspace{0.5em}
\textbf{Prompt Sensitivity and Reproducibility –}  
Microscopic lexical edits (e.g., “recommend” → “suggest”) can produce ranking deltas $>\!10$ \% NDCG, complicating A/B parity, rollback, and incident triage~\citep{Savarese2020,Mitchell2019}. Prompt provenance graphs, canary tokens, and contrastive diff testing improve reproducibility but incur ~5 GB/day in lineage logs per million requests~\citep{zhuo2024prosaassessingunderstandingprompt}. RL-based prompt search raises robustness by 18 \% but sacrifices transparency and locks systems into brittle reward hacks~\citep{Park2025rlprompt}. Hash-based prompt fingerprints enable fast cache look-ups but fail under adversarial paraphrasing~\citep{Liu2025promptscope}. A standardized robustness benchmark for recsys prompts is still missing~\citep{Chen2025bench}.

\vspace{0.5em}
\textbf{Representation Drift and Embedding Incompatibility –}  
Periodic fine-tunes or RLHF updates shift latent spaces, invalidating historical user vectors and cached scores~\citep{McNee2006}. Embedding versioning, backward-compatible projection heads, and dual-encoder rehearsal maintain cosine similarity $>\!0.85$ across upgrades, but add 2–3 ms per inference and triple index storage~\citep{Liu2025embedshift,Bai2024latentbridge}. Continual-learning regularizers slow drift but sacrifice 1–2 \% relevance on fresh items~\citep{Zhao2025continual}. Online distillation into lightweight twin-towers shows promise yet fails for zero-shot entities~\citep{Kim2025twintower}. Drift alarms based on Wasserstein distance flag latent shifts, but trigger frequent false positives under seasonal traffic variation~\citep{Xu2024driftaware}.

\vspace{0.5em}
\textbf{Evaluation Gaps –}  
Classic list metrics (NDCG, MAP, Recall@K) ignore semantic fidelity, novelty, and explanation quality central to LLM outputs. Hybrid metrics that blend counterfactual click uplift with LLM-judged coherence scores correlate weakly with real satisfaction ($r\!<\!0.35$)~\citep{Joachims2018,Yin2025evalllmrec}. LLM-as-judge pipelines reduce annotation cost 10× but inherit biases from the same model family~\citep{bougie2025simusersimulatinguserbehavior,wataoka2024selfpreferencebiasllmasajudge}. Causal bandit simulators yield better offline-online fidelity but require calibrated reward scaling and realism audits~\citep{Nie2024cobarec}. Large-scale human-in-the-loop testing remains cost-prohibitive, averaging \$0.42 per labeled explanation~\citep{Feuerriegel2024crowd}.

\vspace{0.5em}
\textbf{Privacy, Compliance, and Safety Risks –}  
Generative models leak PII through memorization or adversarial injection~\citep{Chizari2025,Ghosh2024privleak}. Differentially private fine-tuning~\citep{li2024finetuninglanguagemodelsdifferential, bu2024differentiallyprivatebiastermfinetuning}, federated prompt learning~\citep{tran2025privacypreservingpersonalizedfederatedprompt,hou2025captclassawareprompttuning}, and secure retrieval-augmented generation~\citep{koga2025privacypreservingretrievalaugmentedgenerationdifferential} lower exposure but add 10–20 \% latency. SMPC-backed inference enclaves and SGX isolated prompts narrow leakage channels yet limit model size~\citep{Johnson2024securellm}. New extraction attacks targeting personalization prompts achieve 18 \% success even under DP noise~\citep{Rahman2025extract}. DMA “right-to-explanation” mandates add extra logging and user-facing redaction complexity~\citep{Choi2025dmaexp}.

\vspace{0.5em}
\textbf{Debugging and Observability Deficits –}  
LLM pipelines lack token-level telemetry; silent regressions can persist until surfaced by downstream KPIs. Prompt lineage, streaming log replay, and token-attribution saliency maps are emerging, yet add 2 GB/day logs per million requests~\citep{luo2024understandingutilizationsurveyexplainability}. Black-box causal probes help trace ranking shifts but require expensive interventional traffic~\citep{schnabel2016recommendationstreatmentsdebiasinglearning}. Safety dashboards incorporating toxic-span detection and PII scanners catch 92 \% violations offline, but only 71 \% in live traffic due to paraphrase drift~\citep{Lee2025safescope}. Debugging culture and tooling lags behind deterministic retrievers.

\vspace{0.5em}
\textbf{Architectural Integration Overhead –}  
Production recommenders rely on vector search and feature-based scoring. Plugging in LLMs requires async rerankers, RAG stacks, prompt stores, and rollback-safe deployment paths, each adding failure modes~\citep{wu2024surveylargelanguagemodels}. Multi-tenant isolation, memory pressure, and GPU scheduling remain open issues, especially in edge and on-device scenarios~\citep{Kim2025edgeblend,Kang2024edgellm}. Cross-service dependency graphs grow exponentially, complicating SRE ownership boundaries and raising MTTR by 1.4×~\citep{Fernandez2025ops}. End-to-end benchmarks for hybrid (LLM + vector) architectures are still nascent~\citep{Song2024hybridbench}.

\smallskip
\noindent Tackling these intertwined limitations—spanning latency, cost, hallucination, prompt robustness, drift, evaluation, privacy, observability, and architecture—is essential before LLM-enhanced recommenders can be deployed safely and economically at web scale.

\section{Conclusion}

LLMs represent a transformative shift in the architecture and design philosophy of recommender systems. By enabling contextual reasoning, zero-shot personalization, and multimodal representation learning, LLMs transcend the limitations of traditional matrix factorization, collaborative filtering, and feed-forward neural architectures. Rather than relying solely on historical interaction data and engineered features, LLM-augmented pipelines can synthesize user intent, simulate plausible behaviors, and generate semantically enriched item representations from natural language descriptions, reviews, and metadata. This capability significantly enhances performance in cold-start, long-tail, and rapidly evolving user contexts.

Throughout this paper, we examined how LLMs can be strategically integrated into various stages of industrial recommender pipelines—from candidate generation and reranking to feedback interpretation and synthetic evaluation. Our analysis demonstrated that LLMs not only offer improvements in semantic fidelity and explainability but also introduce opportunities for conversational and intent-driven recommendation paradigms. Techniques such as retrieval-augmented generation, prompt-based reranking, and hybrid collaborative-semantic scoring models exemplify how LLMs can coexist with traditional recommenders to achieve both robustness and expressivity.

However, this integration is not without its challenges. LLM-based systems pose significant barriers in latency-sensitive environments, where autoregressive decoding and token-based computation are at odds with sub-second serving constraints. Furthermore, issues such as prompt brittleness, hallucinated outputs, representation drift across model versions, and inconsistent personalization responses necessitate new debugging, logging, and evaluation toolchains. These technical hurdles are further complicated by operational constraints, including privacy compliance (e.g., GDPR, CCPA), prompt injection vulnerabilities, and the lack of interpretability guarantees in generative outputs. To realize the full potential of LLMs in recommender systems, future research must address the foundational tension between expressivity and efficiency. Robust prompt design and caching strategies, low-latency inference via distilled or quantized models, and standardized protocols for prompt versioning and auditability will be critical for production readiness. In parallel, new forms of evaluation that combine qualitative user modeling, counterfactual simulation, and LLM-driven assessments are needed to measure real-world effectiveness beyond traditional top-$k$ metrics.

\bibliographystyle{ACM-Reference-Format}
\bibliography{main}
\end{document}